\documentclass[fleqn,10pt]{wlscirep}
\usepackage{subfigure}
\usepackage{amsmath}
\usepackage{graphicx}
\usepackage{authblk}
\usepackage{soul,xcolor}
\usepackage[colorinlistoftodos]{todonotes}

\title{Data-driven Signal Decomposition Approaches: A Comparative Analysis}
\author[1]{Thomas Eriksen}
\author[1,*]{Naveed ur Rehman}
\affil[1]{Department of Electrical and Computer Engineering, Aarhus University, Denmark}

\begin{abstract}
\textcolor{black}{Signal decomposition (SD) approaches aim to decompose non-stationary signals into their constituent amplitude- and frequency-modulated components. This represents an important preprocessing step in many practical signal processing pipelines, providing useful knowledge and insight into the data and relevant underlying system(s) while also facilitating tasks such as noise or artefact removal and feature extraction. The popular SD methods are mostly data-driven, striving to obtain inherent well-behaved signal components without making many prior assumptions on input data. Among those methods include empirical mode decomposition (EMD) and variants, variational mode decomposition (VMD) and variants, synchrosqueezed transform (SST) and variants and sliding singular spectrum analysis (SSA). With the increasing popularity and utility of these methods in wide-ranging application, it is imperative to gain a better understanding and insight into the operation of these algorithms, evaluate their accuracy with and without noise in input data and gauge their sensitivity against algorithmic parameter changes. In this work, we achieve those tasks through extensive experiments involving carefully designed synthetic and real-life signals. Based on our experimental observations, we comment on the pros and cons of the considered SD algorithms as well as highlighting the best practices, in terms of parameter selection, for the their successful operation. The SD algorithms for both single- and multi-channel (multivariate) data fall within the scope of our work. For multivariate signals, we evaluate the performance of the popular algorithms in terms of fulfilling the mode-alignment property, especially in the presence of noise.}

\end{abstract}
\begin{document}
\flushbottom
\maketitle
%
%
\thispagestyle{empty}
\section{Introduction}

Fourier transform (FT) is one of the most widely used signal processing techniques. It enables a dual view of the time domain signal from the `frequency’ perspective. The two representations are exclusive in a sense that the frequency (or Fourier) representation makes no reference to time and vice versa. But while this exclusivity of time and frequency domain representations may seem mathematically powerful, it becomes prohibitive when processing wide-ranging (non-stationary) signals in real-life which exhibit natural entanglements between the two domains.

The above shortcomings of the FT have led to a large and varied body of work, referred to as the T-F analysis, that attempts to view a signal in the joint time-frequency (T-F) domain \cite{Flandrin18}. Here, the key goal is to track the time variations of signal frequencies -- instantaneous frequency. Interestingly, for many real-life signals, their trajectories of instantaneous frequencies appear as a finite number of ribbon-like coherent structures (ridges) in the T-F domain, revealing hidden inner structures of the data. This begs the question: \textit{can we also obtain the time domain representation of these organized structures appearing in the T-F domain?} Indeed, such representations could further improve our understanding of underlying physical processes and facilitate the processing of non-stationary multi-component signals at the level of their constituent modes in the time domain \cite{Daub11}. We will refer to such time-domain representations of multi-component signals as signal decomposition (SD) in the sequel, though in literature those are also referred to as mode retrieval, mode decomposition and signal separation. 

The above tasks of obtaining signal decomposition (SD) and T-F representations for non-stationary data are directly related to this study and therefore it is pertinent to define those precisely. In relation to SD, let the input signal $x(t)$ be a multi-component signal containing finite $K$ number of components or modes i.e.,
\begin{equation}
x(t)=\sum_{k=1}^{K}c_k(t),
\end{equation}
 
\noindent where $c_k(t)=a_k(t)\cos(\phi_k(t))$, for $k=1\ldots K$, denotes a family of amplitude- and frequency-modulated (AM-FM) components of $x(t)$, which needs to be recovered; $a_k(t)$ and $\phi_k(t)$ respectively represent instantaneous amplitude and phase functions of $c_k$. It is important to mention here that the choice of the AM-FM mode extraction model in (1) is governed by the properties of real-life signals, which naturally involve multiple rhythms and oscillations e.g., circadian \cite{Takeda11} and cortical \cite{Wang10} rhythms, heart rate \cite{Lin14}, respiratory variability \cite{Baudin14}, speech and vibration signals \cite{Dybala14}.

Next, the related problem of obtaining T-F representation of $x$ can be written as
\begin{equation}
x(t) \rightarrow \Phi x(t,f),
\end{equation}
\noindent where $\Phi x(t,f)$ represents the energy density of the signal $x(t)$ in the T-F domain. The T-F analysis and signal decomposition have both become indispensable tools in a wide range of problems, with the T-F analysis finding utility in audio and speech processing \cite{Maragos01,Mallat08}, communication \cite{Matz13}, and biomedical engineering \cite{Park13,Saleem19}; whereas signal decomposition is used in vibration analysis \cite{Dybala14}, denoising \cite{Rehman17,Rehman19-den,Naveed20, Naveed21} 
data fusion \cite{Rehman09-fus,Rehman15,Abd15} 
and medical studies \cite{Daub14}, to name a few.

\begin{figure}[!htbp]
\centering
\begin{tabular}{cc}
  \includegraphics[width=.40\textwidth]{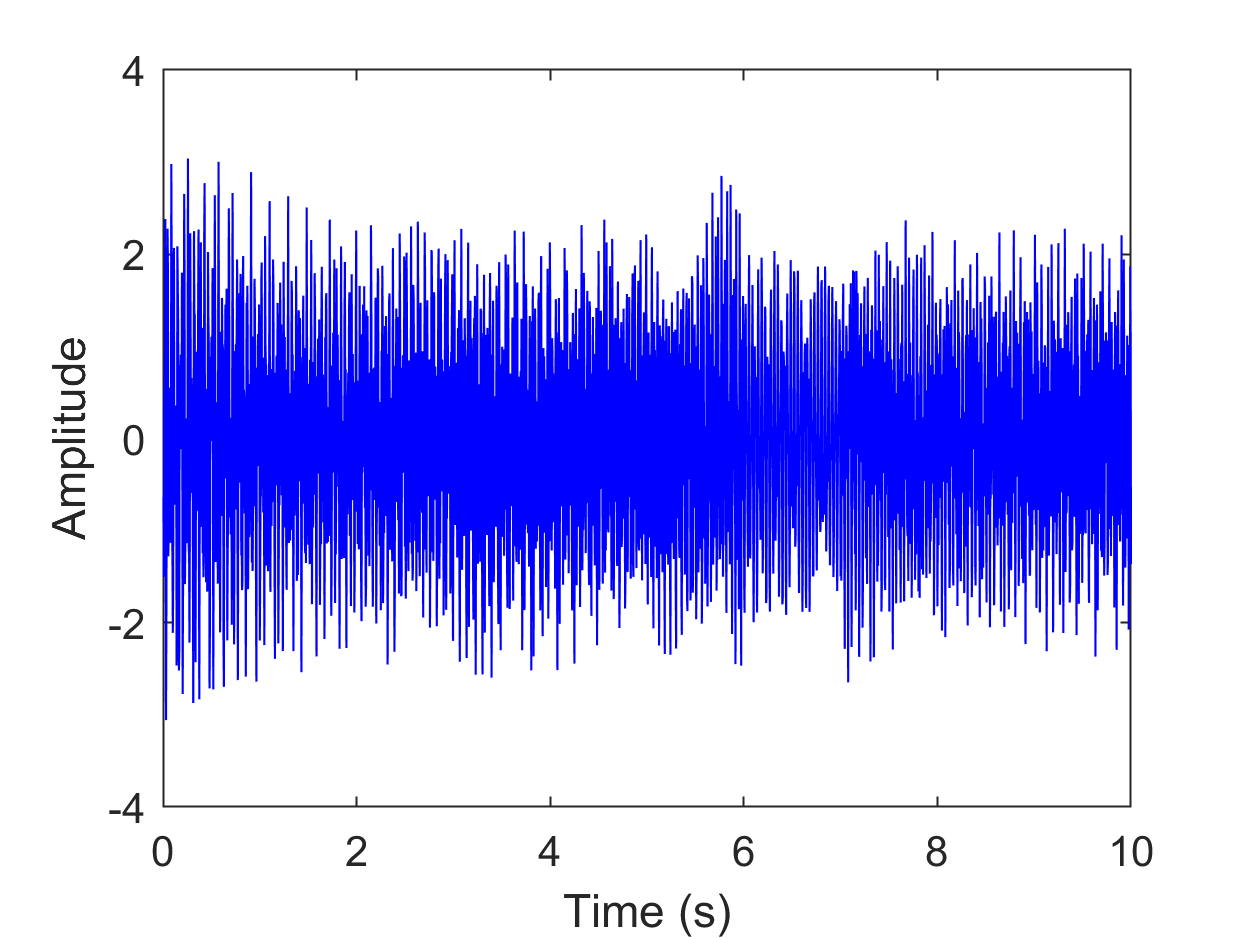} &   \hskip-1.3cm\includegraphics[width=.40\textwidth]{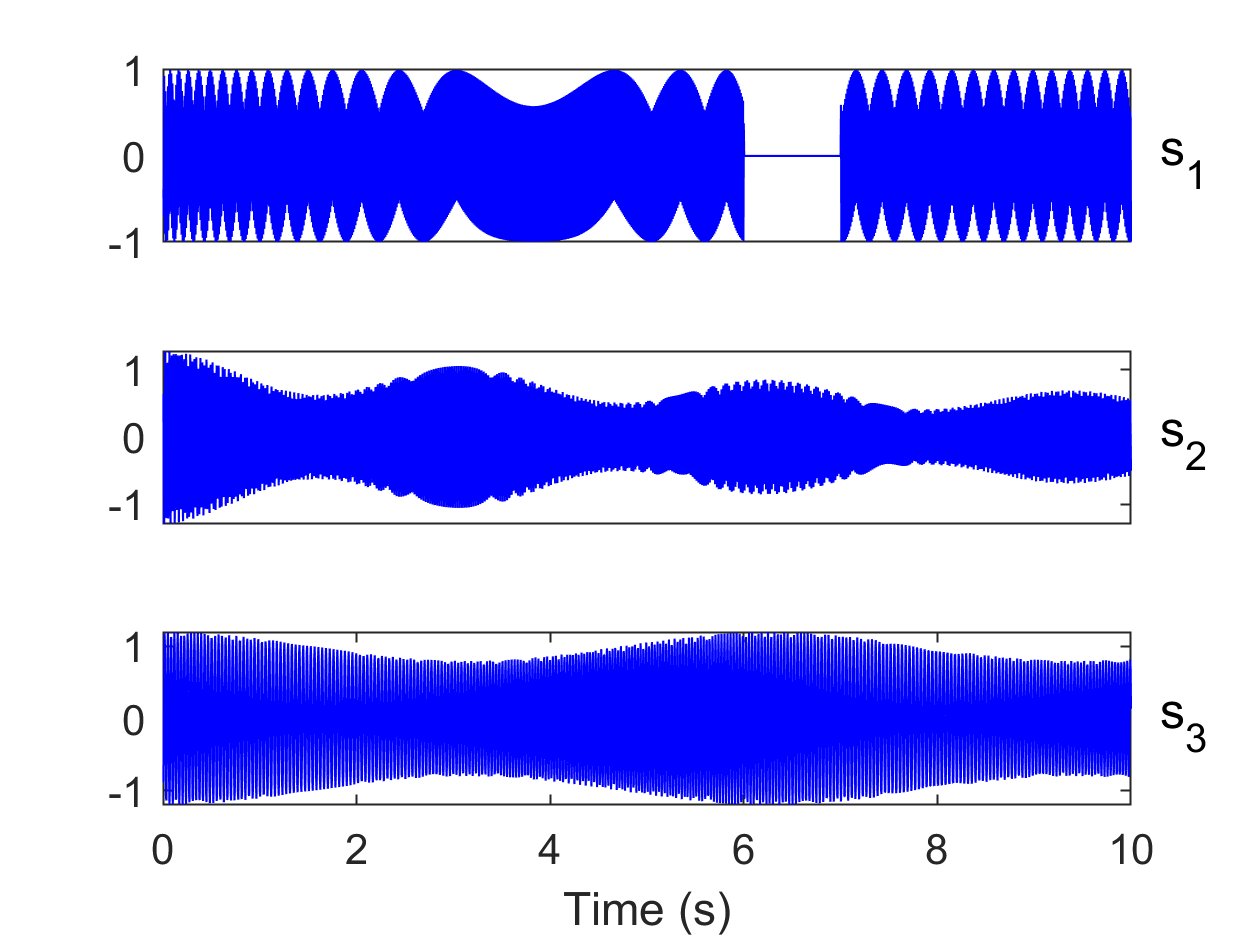} \\
(a)  & \hskip-0.9cm (b)  \\
 \hskip+1cm\includegraphics[width=.475\textwidth]{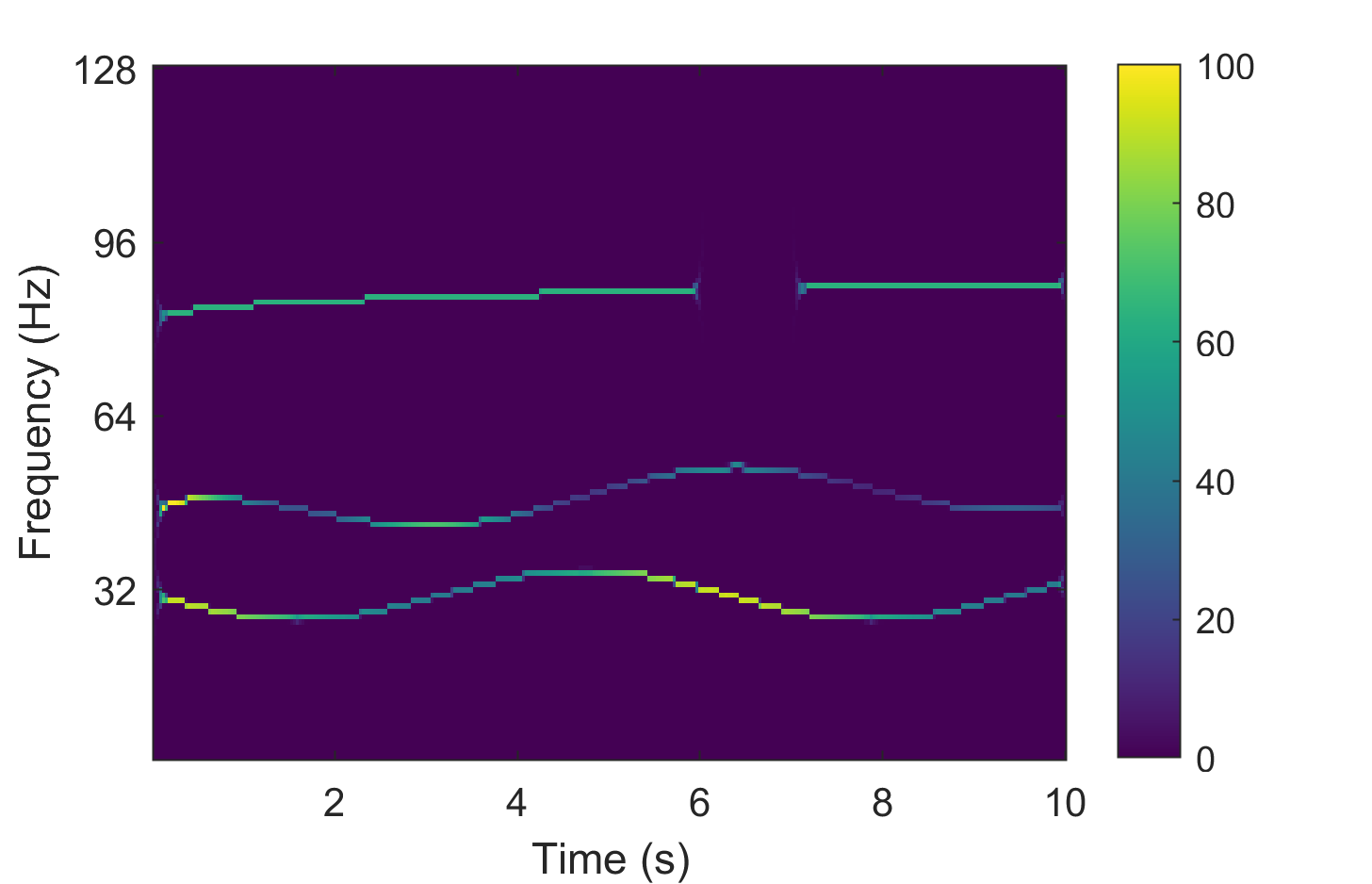} &   \hskip-0.75cm\includegraphics[width=.475\textwidth]{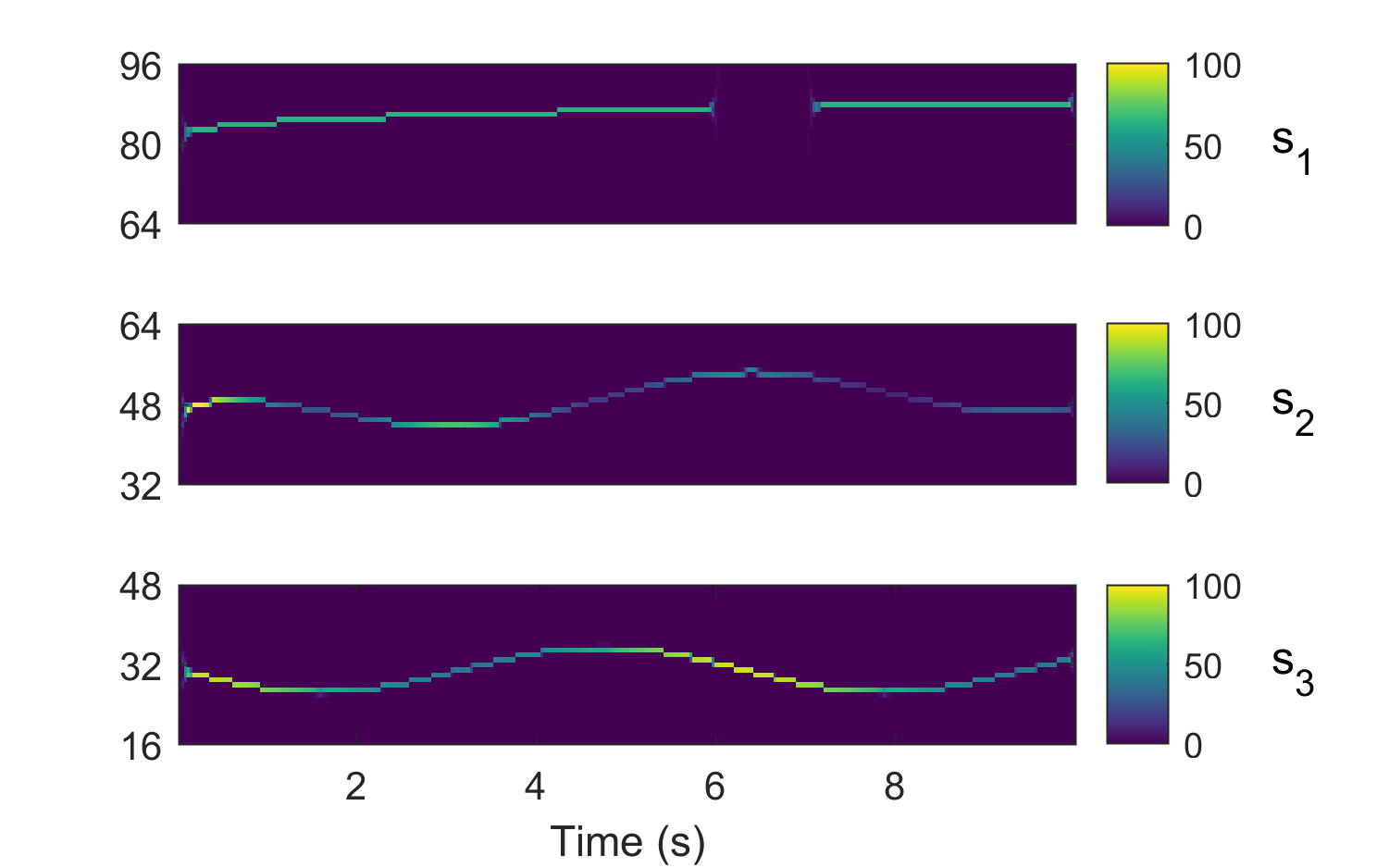} \\
(c) & \hskip-1.3cm (d) \\
\end{tabular}
\caption{Example of SD and T-F representation of a signal. The original composite signal is shown in the time domain in (a) and its separate components, $s_1$, $s_2$ \& $s_3$ are shown in (b). In (c) the T-F representation is shown for the composite signal. In (d), the time-frequency representation of each of the separate components of the original signal is shown.}
\label{figure:SigDecompTF}
\end{figure}

To illustrate the SD and T-F analysis tasks, we show an example of a composite signal $x(t)$ with three components as shown in Fig. \ref{figure:SigDecompTF} (a); each of the three components (modes) $s_1$, $s_2$ and $s_3$ shown separately in Fig. \ref{figure:SigDecompTF} (b). In the second row, the respective T-F plots are shown in Fig. \ref{figure:SigDecompTF} (c) and (d). Given $x(t)$, the goal of SD is to recover its constituent oscillatory components (as given in Fig. \ref{figure:SigDecompTF} (b)) whereas the T-F analysis entails obtaining the T-F plot of the whole signal, shown in  Fig. \ref{figure:SigDecompTF} (c), or the T-F plots of individual components, shown in  Fig. \ref{figure:SigDecompTF} (d).

The earliest attempts at providing a joint T-F representation (2) of non-stationary signals resulted in an intuitive and ad hoc procedure of sliding (or windowed) Fourier transform and Wigner-Villa distribution - a quadratic functional of a signal which can be interpreted as a varying spectral density. The theory of \textit{wavelets} represents another direction in harmonic signal analysis, replacing the notion of `frequency' with `scale' and formalizing the idea of local signal representation at multiple scales.    

The idea of decomposition of a multicomponent signal into its constituent AM-FM components, i.e., the SD problem given in (1), was popularized after the emergence of an empirical data-driven approach known as empirical mode decomposition (EMD) \cite{Huang98}. Despite its inherent flaws that are mainly linked with the empirical nature of the algorithm, the EMD method has found interdisciplinary applications. This led to an explosion of new and improved techniques for SD as well as new application areas where such techniques have been useful. In addition to the EMD, other popular techniques for SD are synchrosqueezed transform (SST) \cite{Thakur13}, variational mode decomposition (VMD) \cite{VMD14} and its extension, named nonlinear chirp mode decomposition (NCMD) \cite{NCMD18} and sliding singular spectrum analysis (SSA) \cite{Fland18}. Following the SD, one way to obtain the T-F representation of a signal from its decomposed components, $c_k(t)$, is through the application of Hilbert transform -  as in the case of EMD, VMD and SSA. Contrarily, SST obtains the (squeezed) T-F representation first, followed by the separation of signal components in the time domain.      

Regardless of their mode of operation, each of the above methods has their own strengths and weaknesses, making those suitable for certain classes of signals and applications. Further, almost all the methods are fairly sensitive to their operational parameters and it is challenging to tune those for optimal results, for a given signal. Finally, the performance of all methods degrades, albeit to varying degrees, under noisy data. Given this, it is of considerable interest to gain better understanding of the strengths and weaknesses of the popular SD approaches and to evaluate their performance in terms of the i) accuracy of obtaining the constituent AM-FM oscillatory components; ii) robustness against noise; iii) sensitivity to changes in the algorithmic parameters. 

To this end, we investigate and compare the performance of the popular SD approaches in the context of their accuracy of decomposition, their ability to operate under noisy data and their sensitivity to algorithmic parameter changes. We also make suggestions regarding the optimal choice of algorithmic parameters of the SD approaches. Our observations and suggestions are based on the empirical results of carefully designed experiments using synthetic signals as well as real-life data. We consider and examine tbhe performance of both single-channel (or univariate) SD approaches and their multi-channel (multivariate) extensions in this study. Finally, it is emphasized that dozens if not hundreds of data-driven SD techniques have emerged over the last two decades and it is not within the scope of this work to assess or even review all those; instead, we focus only on a few largely popular classes of SD and T-F approaches in this work.

The article is organized as follows: Section 2 gives an overview of the state-of-the-art in SD along with some technical details of the methods considered in this study. Section 3 presents experiments and related observations regarding the accuracy of the SD approaches, while section 4 examines the noise robustness of the considered methods. Section 5 evaluates the performance of the multivariate extensions of the popular SD methods in the presence of noise. Section 3-5 also examine the sensitivity of the SD methods to parameter changes. The article is concluded with an overall performance assessment of the considered approaches along with related discussion.

\section{Review of the Popular SD Approaches}

In this section, we briefly review the well-established and popular SD techniques which have been considered for evaluation in this work. The goal is to give a general description of each technique without going into algorithmic details, though we do mention important algorithmic parameters that will be relevant in the subsequent analysis. Further details of the relevant algorithms can be found in the provided references.  

\subsection{Empirical Mode Decomposition (EMD) and variants} 
EMD is the first truly data-driven method for signal decomposition and T-F analysis \cite{Huang98,Huang09}. It operates through a sifting process that iteratively extracts inherent oscillations from input data based on signal extrema and their interpolation. 
The downside of EMD is that it lacks a rigorous mathematical framework, which poses problems in guaranteeing its correctness, stability and performance assessment in general. Efforts to overcome those difficulties include convex approximations of EMD \cite{Meignen07, Pustelnik14}, local iterative filtering approaches \cite{Lin09, Cicone16}, sparse models to obtain data-driven modes \cite{Hou11, Hou16}, and noise-assisted approaches \cite{Wu09,Rehman13-AADA, Lang20}.

In this article, we consider the original EMD method by Huang \textit{et. al} \cite{Huang98}. The important operational choices within the EMD method are: i) the extrema interpolation scheme; ii) the stopping criterion for the intrinsic mode functions. We employ the well-established cubic spline approach for extrema interpolation and use the stopping criterion, introduced in \cite{EMD-alg}, that is based on two threshold values.  

\subsection{Synchrosqueezed Transform (SST) and variants} 
Founded on strong mathematical footing, SST is a powerful tool for nonstationary signal decomposition and T-F analysis \cite{Daub96,Daub11,SST_SPM13}. The method operates by sharpening the T-F representation of a signal by using a synchrosqueezing operator, followed by the T-F ridge extraction and component retrieval. Several variants of SST have emerged in the last decade, including its adaptation to STFT \cite{Thakur11, Thakur13}, S-transform \cite{Huang16} and multitapered transform \cite{daub15}; higher order variants \cite{Oberlin15, Bahera18, Pham17, Yang18}; improved signal separation operators \cite{Chui16, Li20}; and ridge extraction techniques \cite{Meignen17, Laurent20}. 

The performance of SD task via SST relies heavily on i) the ridge extraction technique; ii) the synchrosqueezed T-F (or time-scale) representation of input data. To create the initial T-F (or time-scale) representation, appropriate mother wavelet function and accompanying parameters must be chosen. An optimization based heuristic approach \cite{Thakur13} is typically used for ridge extraction within SST. The approach requires the tuning of two crucial parameters for its operation: i) the initial bandwidth $\omega_i$; ii) the maximum step size $\alpha$ which restricts the maximum change in center frequency along the time-axis.

\subsection{Variational Mode Decomposition (VMD) and variants}
VMD poses the signal decomposition problem within the mathematical framework of variational convex optimization \cite{VMD14}. Despite its popularity, VMD is applicable only to signals containing bandlimited components; a recent work, titled nonlinear chirp mode decomposition (NCMD), aims to address that problem but suffers from serious convergence issues \cite{NCMD18}.

Both VMD and NCMD will be considered in our analysis. The important parameters affecting the operation of the VMD method are: i) the number of components $K$ to be separated; ii) the bandwidth parameter $\alpha$; and iii) $\tau$, an internal parameter enforcing the exact reconstruction via Lagrangian multiplier. The precision of reconstruction can be changed by altering $\tau$. For NCMD, in addition to $K$, $\alpha$, and $\tau$, tuning the increment parameter $\mu$ is crucial as it affects the estimation of change in mode frequencies. More information about the parameters of VMD \cite{VMD14} and NCMD \cite{NCMD18} can be found in the relevant articles.

\subsection{Source Separation based Approaches} Source separation methods based on simple generative models bear close resemblance with non-stationary signal decomposition techniques due to their shared data-driven flavour \cite{Comon94, Lee99, Smaragdis14}. Yet, this connection between the two classes of methods has largely remained unexplored. Few notable exceptions, however, exist including singular spectrum analysis \cite{Broomhead86}, sliding SSA (SSA) \cite{Fland18} and low-rank T-F synthesis (LRTFS) approach \cite{LTFRS18}. 

In this work, SSA \cite{Fland18} is used as one of the assessed approaches. An important parameter that requires tuning within SSA is the embedding dimension $L$. Moreover, the desired number of classes/clusters is another user-defined parameter that affects the performance of the algorithm. 

\subsection{Multivariate Extensions of the SD Approaches}
Specialized algorithms for multivariate data have become crucial in modern applications, owing to recent developments in sensing technology. To this end, the notable multivariate extensions of data-driven NSP methods include those for EMD \cite{Rehman09,Rehman10,Rehman11,Rehman13}, local iterative filtering \cite{Cicone19}, VMD \cite{Rehman19,Chen20}, SST \cite{Ali15}, empirical wavelet transform \cite{MEWT18} and sliding SSA \cite{Jain20}. 

In this article, we will evaluate the performance of multivariate extensions of EMD, named MEMD \cite{Rehman13}, and VMD, known as MVMD \cite{Rehman19}. Importantly, the algorithmic parameters of both the methods are mostly the same as their univariate extensions. 

\section{Accuracy of Signal Decomposition }
\label{Section3_accuracy}
We evaluate the performance of the popular SD approaches (EMD, VMD, SST, VNCMD and SSA) in terms of accurately decomposing the constituent AM-FM oscillatory components of a multi-component signal. The accuracy of such methods is inherently dependent on the (complexity of) input signal and so careful consideration has been made while choosing the test signals. In particular, we design two synthetic signals comprising of narrow- and wide-band components respectively. Further, a real-life biomedical electroencephalogram (EEG) signal is also used to evaluate the performance of the methods. 

To compare the accuracy of the SD methods quantitatively, we use a metric called quality of reconstruction factor (QRF) \cite{Fland18}. QRF of extracted signal component $\hat{s}$, relative to the reference `true' component $s$, is defined as 

\begin{equation}
\mathrm{QRF}(\hat{s}, s)=20 \times \log _{10}\left(\frac{\|s\|}{\|s-\hat{s}\|}\right).
\end{equation}

Note from (3) that greater (smaller) the error between the reconstructed signal component $\hat{s}$ and the `true' component $s$, the smaller (greater) the value of QRF. Thus, higher values of QRF imply that the SD has been performed accurately. Clearly, the QRF metric can only be used when `true' signal components (or ground truth) is available. 

In this and the remaining sections, rather than showing the time-series plots of the decomposed components, we will show the T-F representation of the individual components obtained by using the Hilbert transform. This provides a more convenient and compact visual representation of the decomposed components.  

\subsection{Case Study 1: Narrow-band Signal}
\label{acc_sig1}
\begin{figure}[!htb]
\centering
   \includegraphics[width=1.\linewidth]{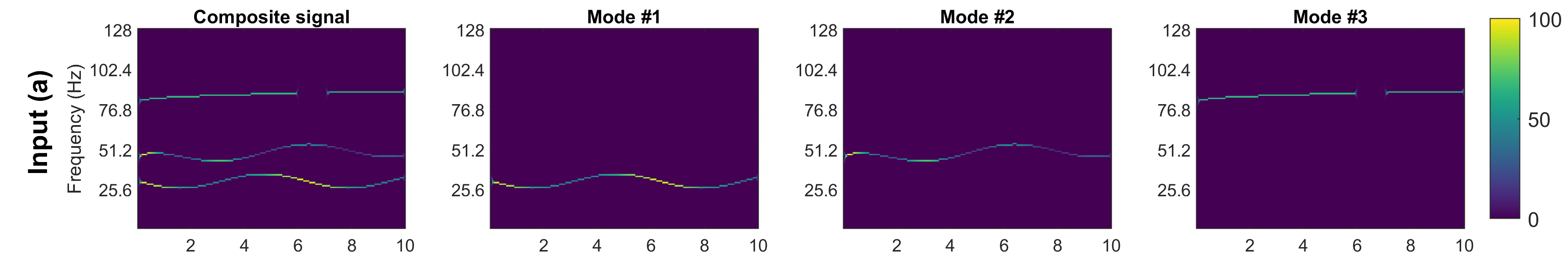}
   \includegraphics[width=1.\linewidth]{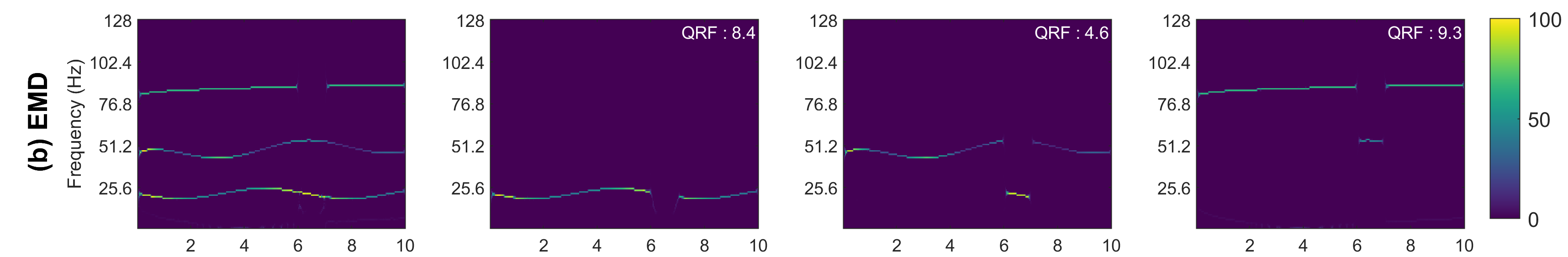}
   \includegraphics[width=1.\linewidth]{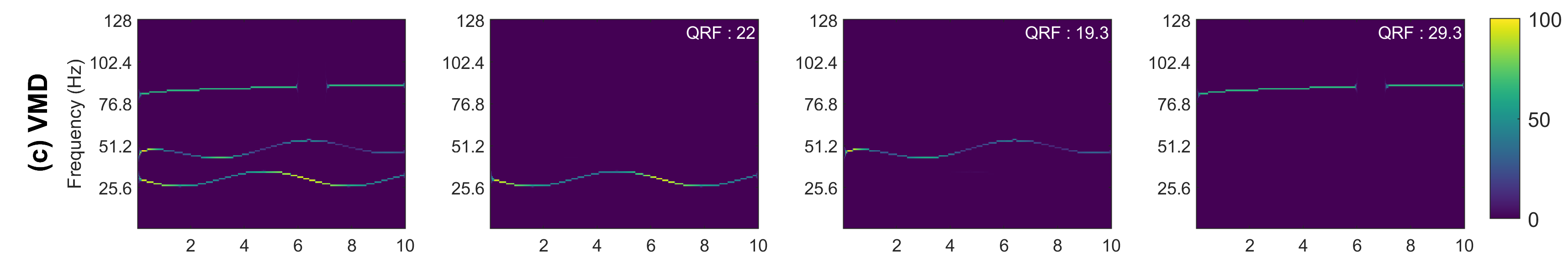}
   \includegraphics[width=1.\linewidth]{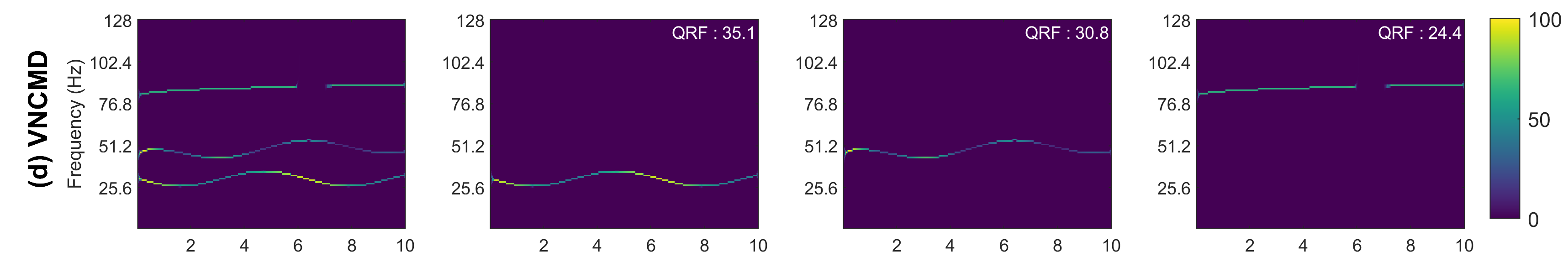}
   \includegraphics[width=1.\linewidth]{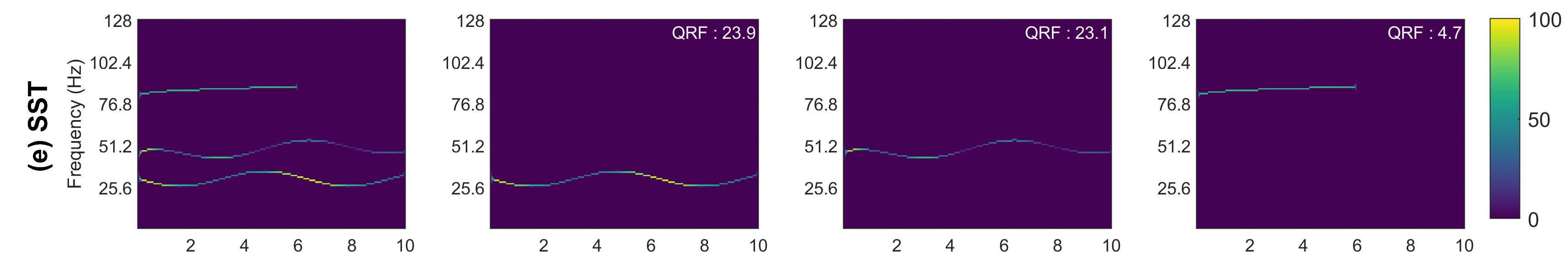}
   \includegraphics[width=1.\linewidth]{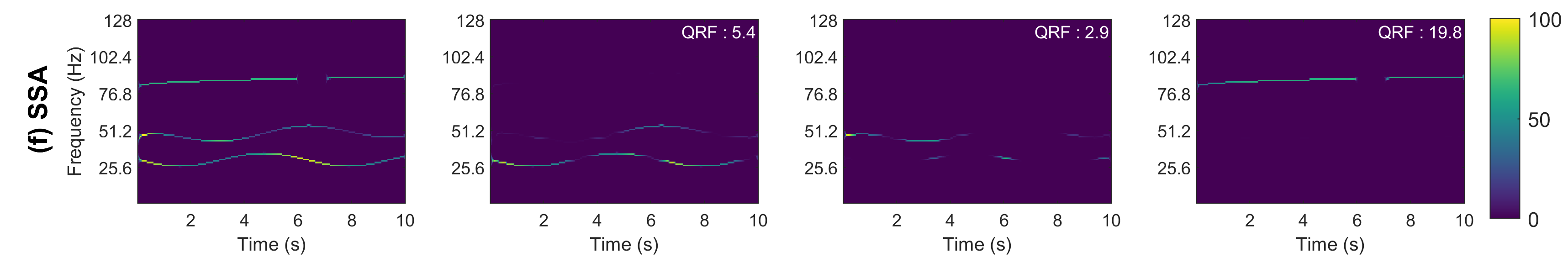}
\caption{Performance evaluation of SD and T-F approaches on the narrow-band synthetic signal. Plot (a) shows the original input signal with each of its constituent components shown in the same row. Similarly this is done for each method, EMD, VMD, VNCMD, SST and SSA in plots (b-f), where the reconstructed signal is shown together with the respective extracted modes. QRF is calculated according to the corresponding original sub-signal and shown in the upper right corner of each T-F plot.}
\label{sec3_syntheticsignal1}
\end{figure}

The first input signal was created synthetically and consisted of 3 components or subs-signals. Each component was an AM-FM signal with narrow-band characteristics. This signal is a modified version of a signal employed in another study \cite{Thakur13}, with a discontinuity in the T-F plane introduced in one of the components to increase the complexity of the signal. The 3 components of the composite signal are defined as, $ s_{11} = (1 + 0.2\cos(t))\cos(30\pi(2t+0.3\cos(t)))$, $ s_{12} = (1 + 0.3\cos(2t))\mathrm{e}^{-\frac{t}{15}}\cos(30\pi(2.4t+0.5t^{1.2}+0.3\sin(t)))$, and $ s_{13} = \cos(30\pi(5.3t+0.2t.^{1.3}))$, which are added together to form the synthetic signal 1, $s_1 = s_{11} + s_{12} + s_{13}$. The signal has a duration of 10 seconds with a sampling rate of 256 Hz. The ideal T-F representation of $s_1$ along with each of its constituent modes is shown in Figure \ref{sec3_syntheticsignal1} (a); notice the T-F discontinuity in mode \#3.

The accuracy of the SD methods is evaluated on the synthetic signal ($s_1$) through the decomposed components and T-F plots of $s_1$ along with the corresponding QRF values. The T-F plots obtained for each method and their respective QRF values are shown in Figure \ref{sec3_syntheticsignal1} (b-f). The input signal is depicted in (a) with its three components shown on its right. The color bar shown to the right of the figure applies both to the composite signal along with its 3 components. These results were obtained by manually tuning the algorithmic parameters to their optimal values for each method. The impact of changing the algorithmic parameters on the performance of different methods is discussed in section \ref{parameterSensitivity}.

Figure \ref{sec3_syntheticsignal1} and the corresponding QRF values show that the EMD (a), SST (e) and SSA (f) are the worst performing methods for the narrow-band signal. For EMD, we clearly see mode-mixing within all of the extracted modes, resulting in a poor signal decomposition and hence lower values of QRF for all the 3 components. The discontinuity in the original component-3 results in the mode-3 'closing' this gap by taking signal components from mode 2. This indicates that the EMD does not perform well in the presence of T-F discontinuity.

In the case of SST, we see mixed results across the different components. The first two components $s_{11}$ and $s_{12}$ are recovered very well by the SST method as illustrated by the high QRF values. The third component $s_{13}$ was not recovered completely as a result of the ridge extraction being unable to handle the T-F discontinuity. However, it is important to note that these results were obtained by setting the number of extracted components $K=3$. For $K=4$, the SST method would obtain accurate T-F representation of the signal, with the fourth extracted mode containing the missing part of the $s_{13}$. 

SSA was performed with $L = 110$ embedding dimensions, with the results being widely unstable. Small alterations in the the amount of dimensions resulted in widely different results, with $L=110$ producing the best results. Similarly to EMD, we see mode mixing in SSA decomposition. This however only occurs in the first 2 components due to those being too close. By shifting the first component down in the frequency domain, we observed the improved results with the SSA. The $s_{13}$ was recovered well by the SSA method as illustrated by the high QRF value. 

Figure \ref{sec3_syntheticsignal1}(d) shows that the VNCMD performs especially well in terms of decomposing/reconstructing $s_{11}$ and $s_{12}$, delivering the best decomposition out of all the methods. This of course is highlighted by the corresponding QRF values. The method also reconstructed $s_{13}$ fairly well with the corresponding $QRF=24.4dB$. Despite its impressive performance, the VNCMD method comes with the caveat that the starting frequency of each component must be initialized close to the actual `ground truth'. For instance, in this experiment, the starting frequencies where chosen to be 30 Hz, 50 Hz and 85 Hz respectively to obtain these results. Choosing slightly different values for the starting center frequencies of the 3 modes resulted in sub-optimal results. Overall, we found the VNCMD to be  heavily reliant or sensitive to the algorithmic parameter changes - explained further in section \ref{parameterSensitivity}.

Finally, VMD offered the best performance for the narrow band signal as depicted in Figure \ref{sec3_syntheticsignal1}(c). The T-F plots and the corresponding QRF values for all 3 components confirm this observation. Moreover, the method was overall robust to the parameter changes e.g., the above results were unchanged despite varying the parameter $\alpha$ in the range of 30-1000.  

\subsection{Case Study 2: Wide-band Signal}
\label{acc_sig2}
\begin{figure}[h!]
\centering
   \includegraphics[width=0.975\linewidth]{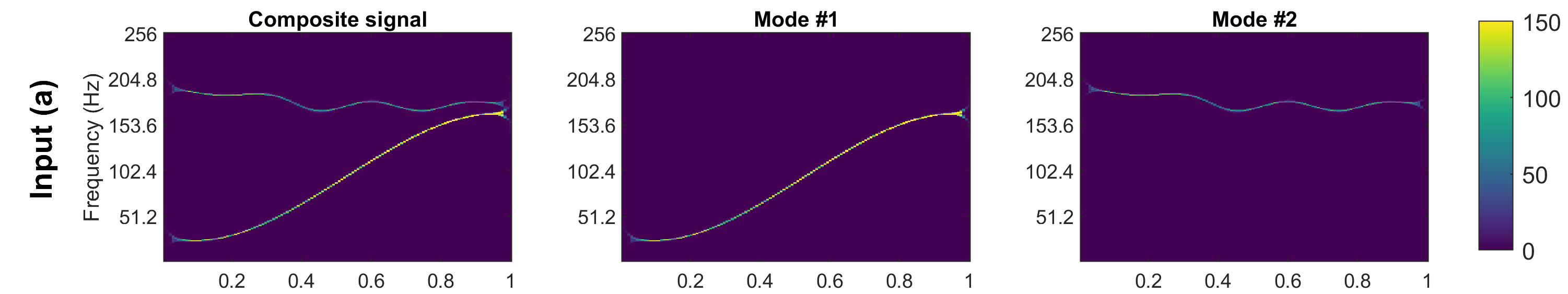}
   \includegraphics[width=0.975\linewidth]{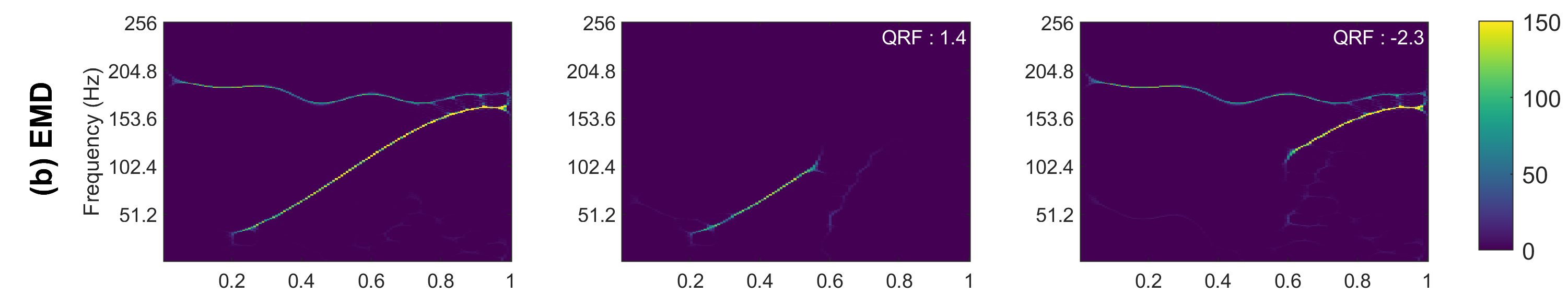}
   \includegraphics[width=0.975\linewidth]{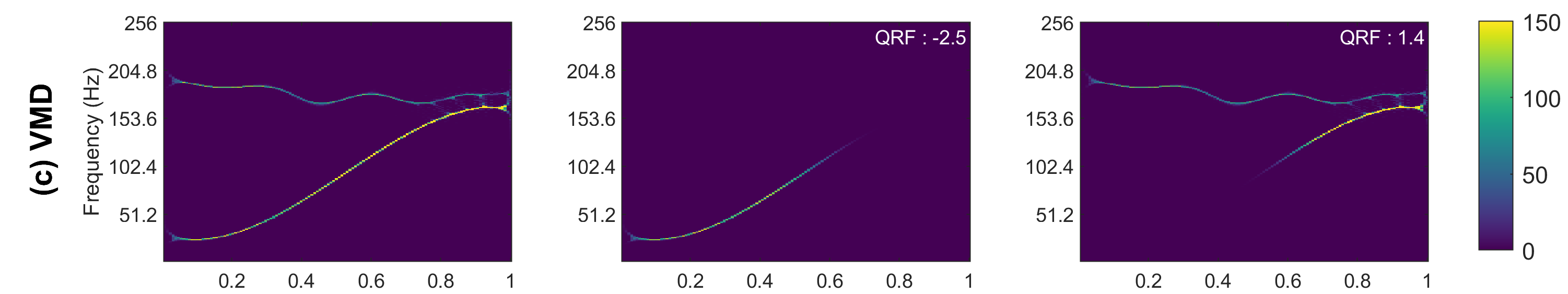}
   \includegraphics[width=0.975\linewidth]{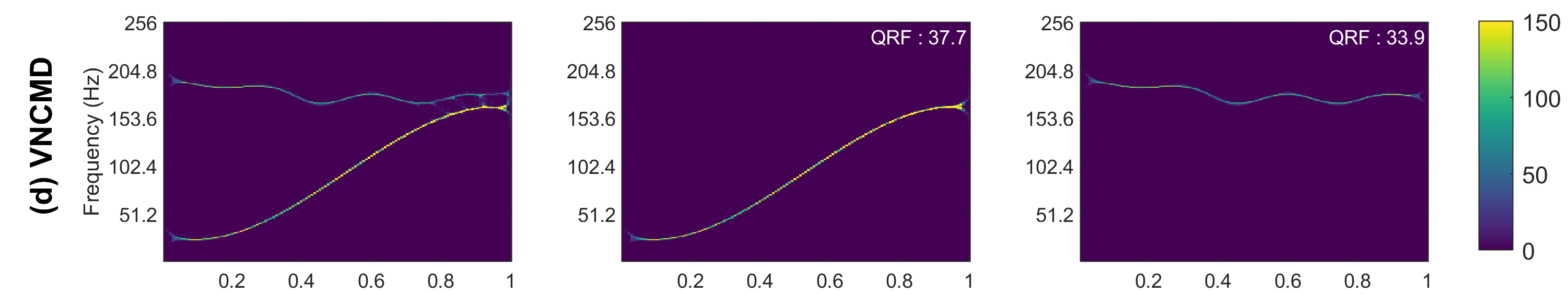}
   \includegraphics[width=0.975\linewidth]{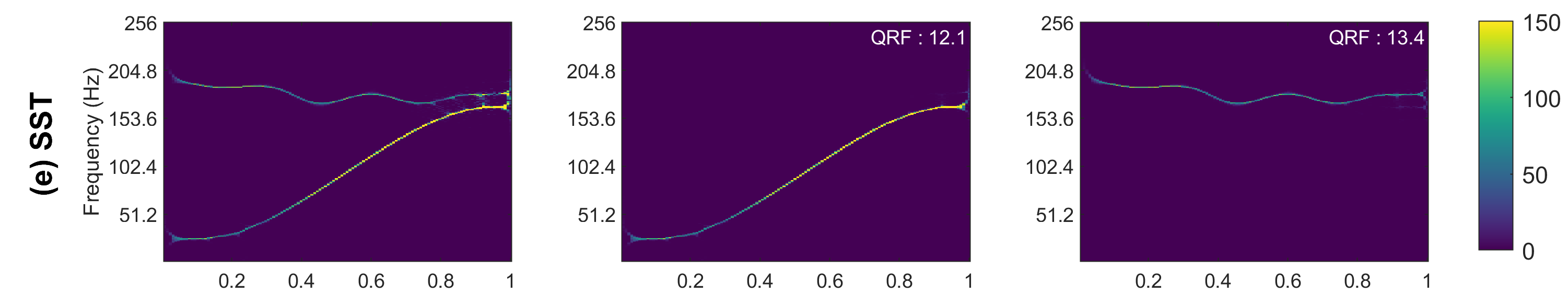}
   \includegraphics[width=0.975\linewidth]{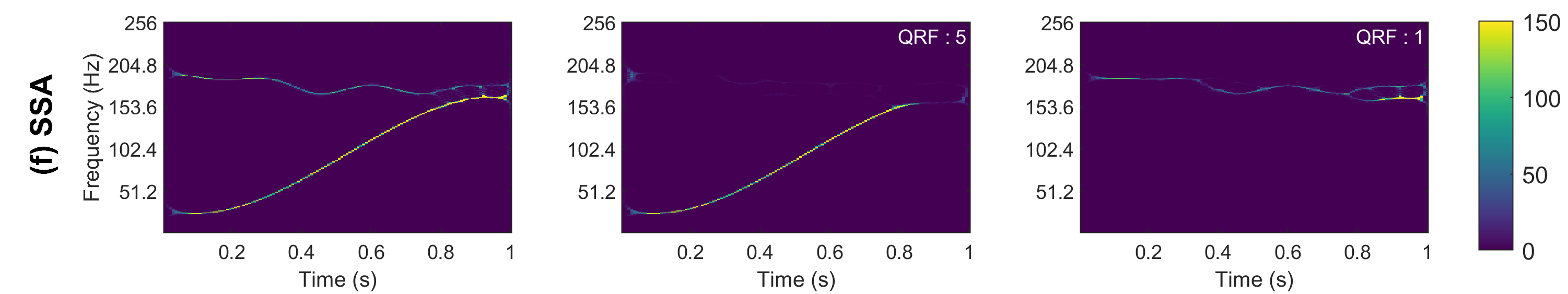}
   
\caption{Performance evaluation of SD and T-F approaches on the wide-band synthetic signal. Plot (a) shows the original signal with each of its different modes separately. Similarly this is done for each method, EMD, VMD, VNCMD, SST and SSA in plots (b-f), where the reconstructed signal is shown together with the respective modes that are extracted. QRF is calculated according to the corresponding original signal and shown in the upper right corner of each TF plot.}
\label{sec3_syntheticsignal2}
\end{figure}

The second test signal was a synthetic signal consisting of 2 components: a wide-band chirp signal within a frequency range of approximately 50-150 Hz and a narrow-band signal generated from a sinusoid modulated with a Gaussian-smoothed Brownian motion. That gives a signal which does not exhibit purely sinusoidal changes over time but instead varies randomly. The wide-band component is given by $s_{21} = e^{0.8t} cos(1.1\pi (0.8+50t-100t^2+416t^3-200t^4))$. The narrow-band signal is generated through Brownian motion with a drift rate $\mu=-0.1$ and a volatility rate $\sigma=0.1$. The resulting signal is smoothed with a Gaussian filter and then used to modulate a sinusoid, resulting in the component $s_{22}$. The wide-band synthetic signal used in this study is generated by adding the two components, i.e.,  $s_2 = s_{21} + s_{22}$. The T-F representation of $s_{2}$ along with its two constituent modes in the same row is shown in figure \ref{sec3_syntheticsignal2} (a). This signal is challenging from two aspects: i)  due to its wide-band nature; ii) because of the closeness of its two components at the end of the signal. 
The signal has a duration of 1 second with a sampling rate of 512 Hz. 

Similar to the analysis of $s_1$, we will evaluate the accuracy of the SD methods on $s_2$ both qualitatively and quantitatively via the T-F plots of the decomposed components and the corresponding QRF measure respectively. In our experiments, various algorithmic parameters of different methods were optimized for $s_2$. 
The T-F plots obtained for each of method along with their respective QRF values are shown in Figure \ref{sec3_syntheticsignal2} (b-f). The original signal is depicted in (a) with its respective components shown in the middle and right columns. 

We start our analysis with the performance of the EMD algorithm: it performed poorly on $s_2$ as reflected in both its T-F plots and the corresponding QRF values. In particular, the wide-band component $s_{21}$ was split into two components with the narrow-band Brownian component being part of the second component. This was expected since by design EMD is suitable for the decomposition of signals with multiple narrow-band components. It is important to note that EMD introduced unwanted artifacts (around $t=0.6 s$ in both modes) which were not part of the original signal. 

Like EMD, the VMD also performed poorly on this wide-band signal. Again, the original wide-band component was decomposed in two separate modes by the VMD. Each decomposed mode therefore exhibits a poor QRF value even though the overall reconstructed signal is close to the original. Unlike EMD though, there were no unwanted artefacts introduced by the VMD algorithm in its decomposed components, as shown in the Figure \ref{sec3_syntheticsignal2} (c). 

The SSA also produced sub-optimal results in terms of component decomposition or construction, as depicted by the low QRF values for both the extracted components. While it appears that the T-F plots of the decomposed components are close to the original, closer inspection reveal that there is mixing of information in both modes. 

The VNCMD performed very well on the wide-band signal but with the caveat that its performance was very unpredictable with respect to initial conditions of the algorithm and its parameter setting. Initial frequency estimates for each decomposed component had to be close to the original ones for the method to work properly. Further, even small changes to the parameter and initial conditions considerably altered the final results. This highlights the serious convergence issues within the VNCMD method. Such limitations of the VNCMD method become pronounced in the presence of noise, as will be noticed in section \ref{Section4}. 

Finally, the SST produced overall good results as shown in Figure \ref{sec3_syntheticsignal2} (e). Despite the lower QRF values as compared to VNCMD, the SST was considerably more stable to parameter changes. This is in contrast to the EMD and VMD methods which are by design applicable to the signals containing narrow-band components.

\subsection{Case Study 3: Electroencephalogram (EEG) Signal}
\begin{figure}[h!]
\centering
\begin{tabular}{cc}
  \includegraphics[width=.375\textwidth]{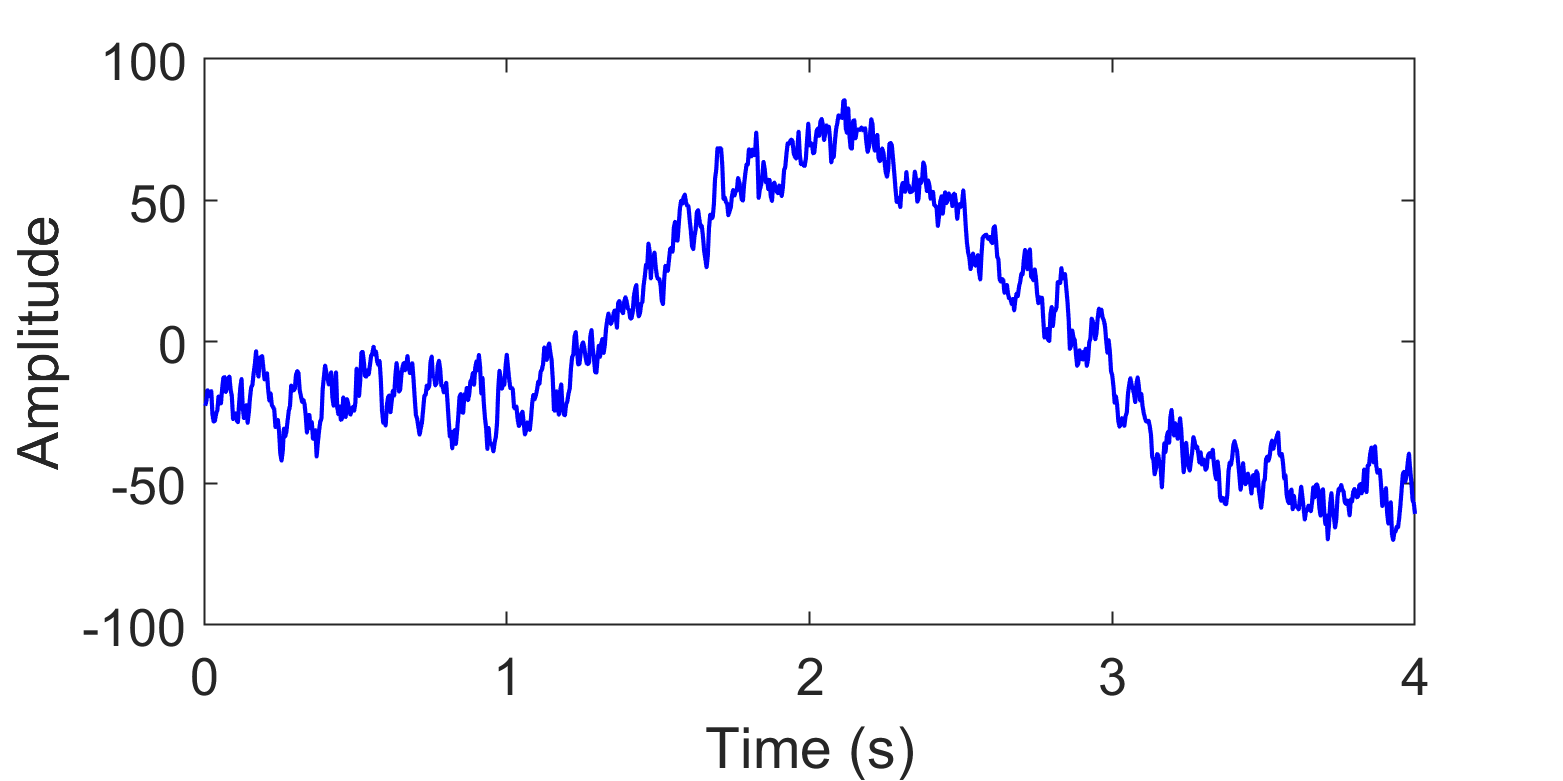} &   \hskip-1.0cm\includegraphics[width=.375\textwidth]{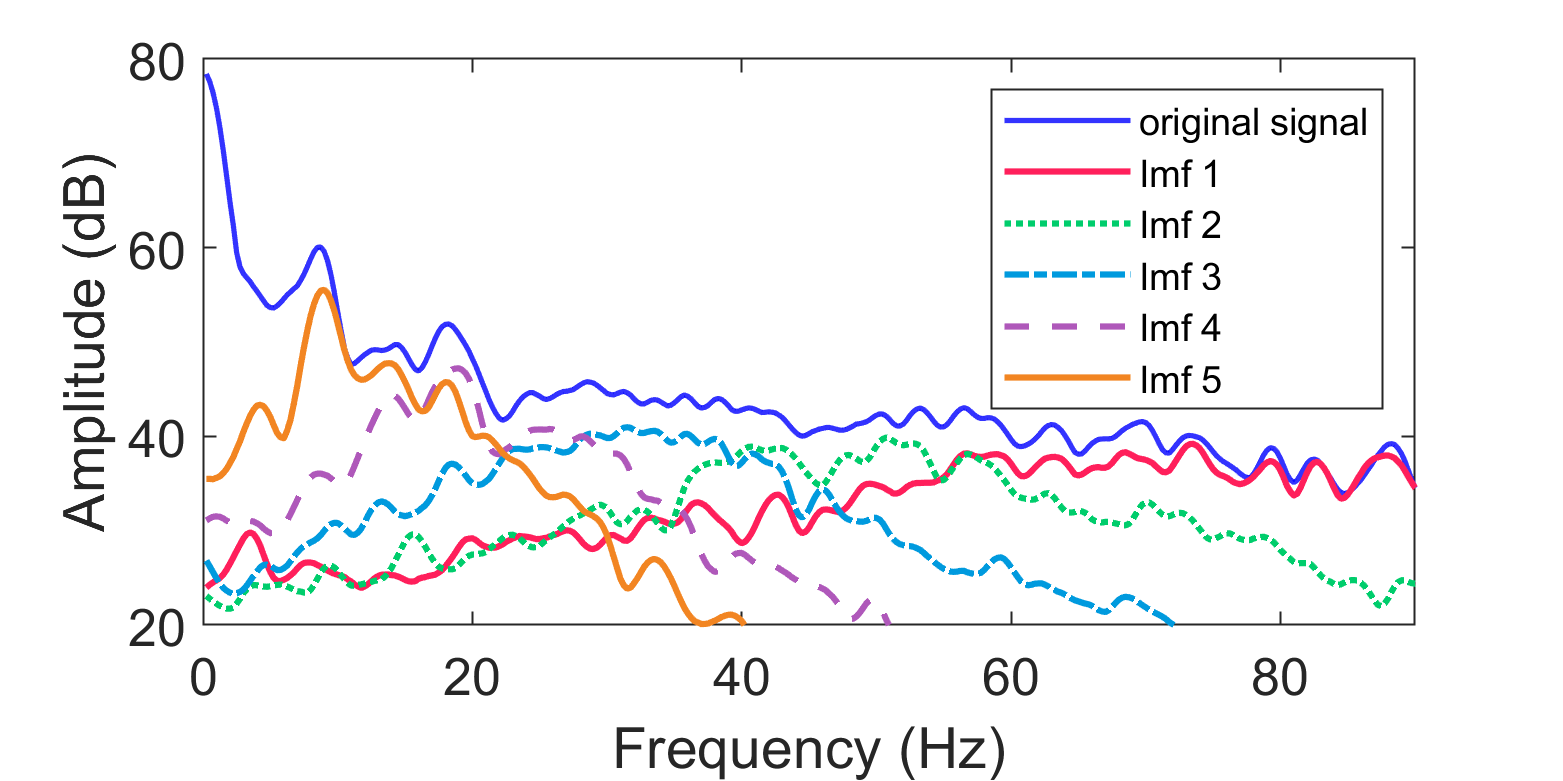} \\
(a) EEG signal time series & \hskip-1.0cm (b) EMD  \\
 \includegraphics[width=.375\textwidth]{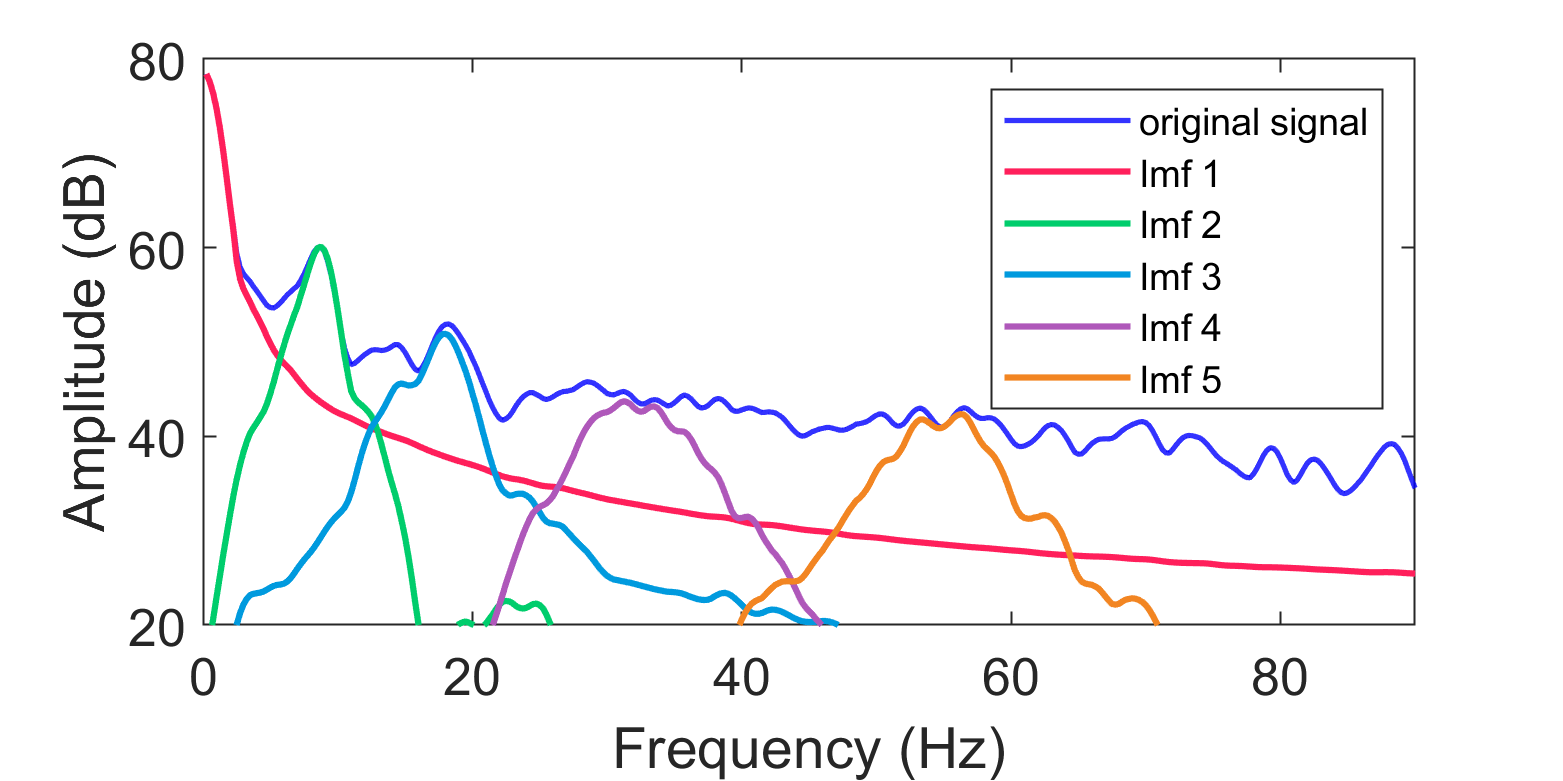} &   \hskip-1.0cm\includegraphics[width=.375\textwidth]{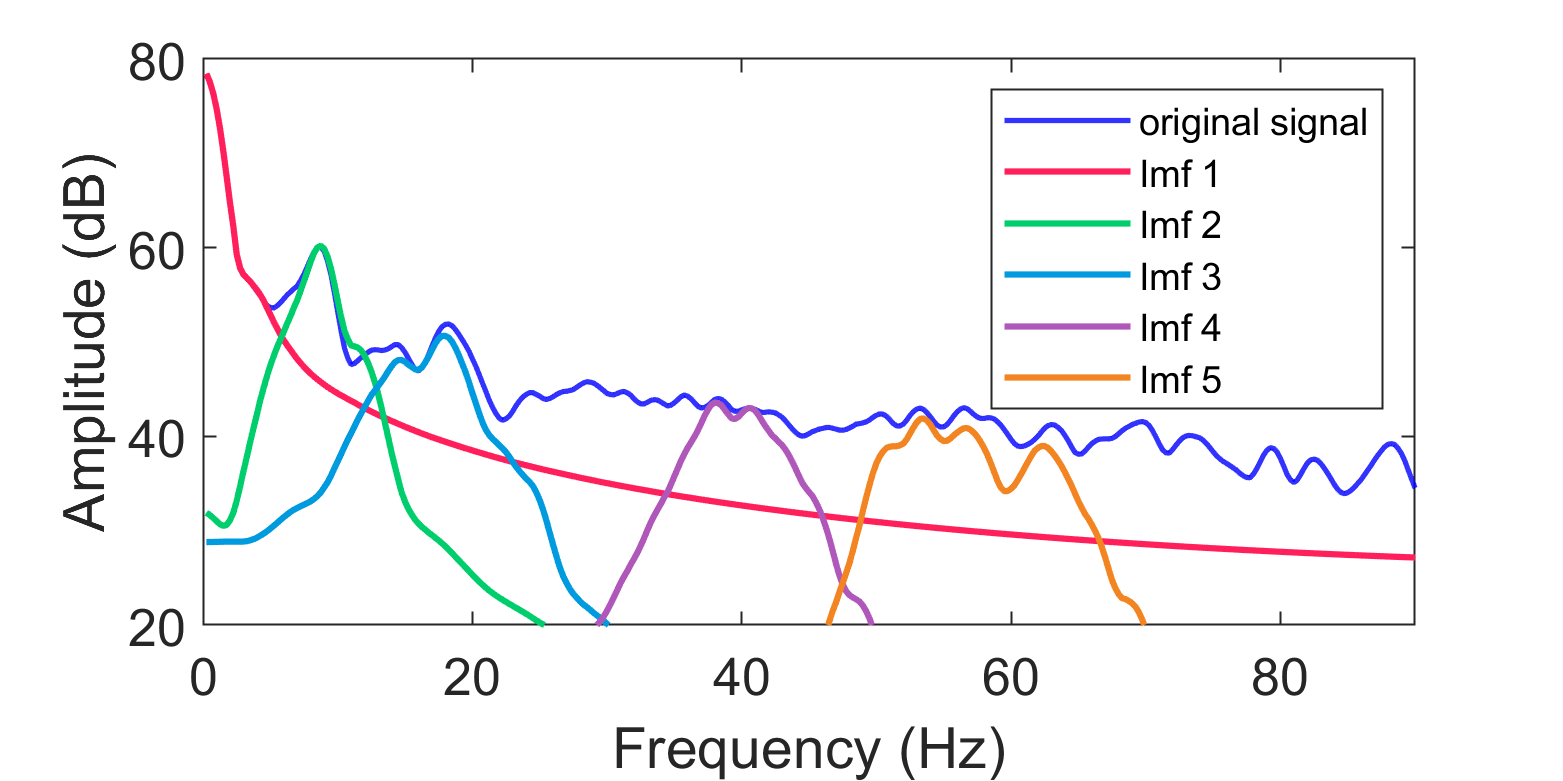} \\
(c) VMD & \hskip-1.0cm (d) VNCMD \\
 \includegraphics[width=.375\textwidth]{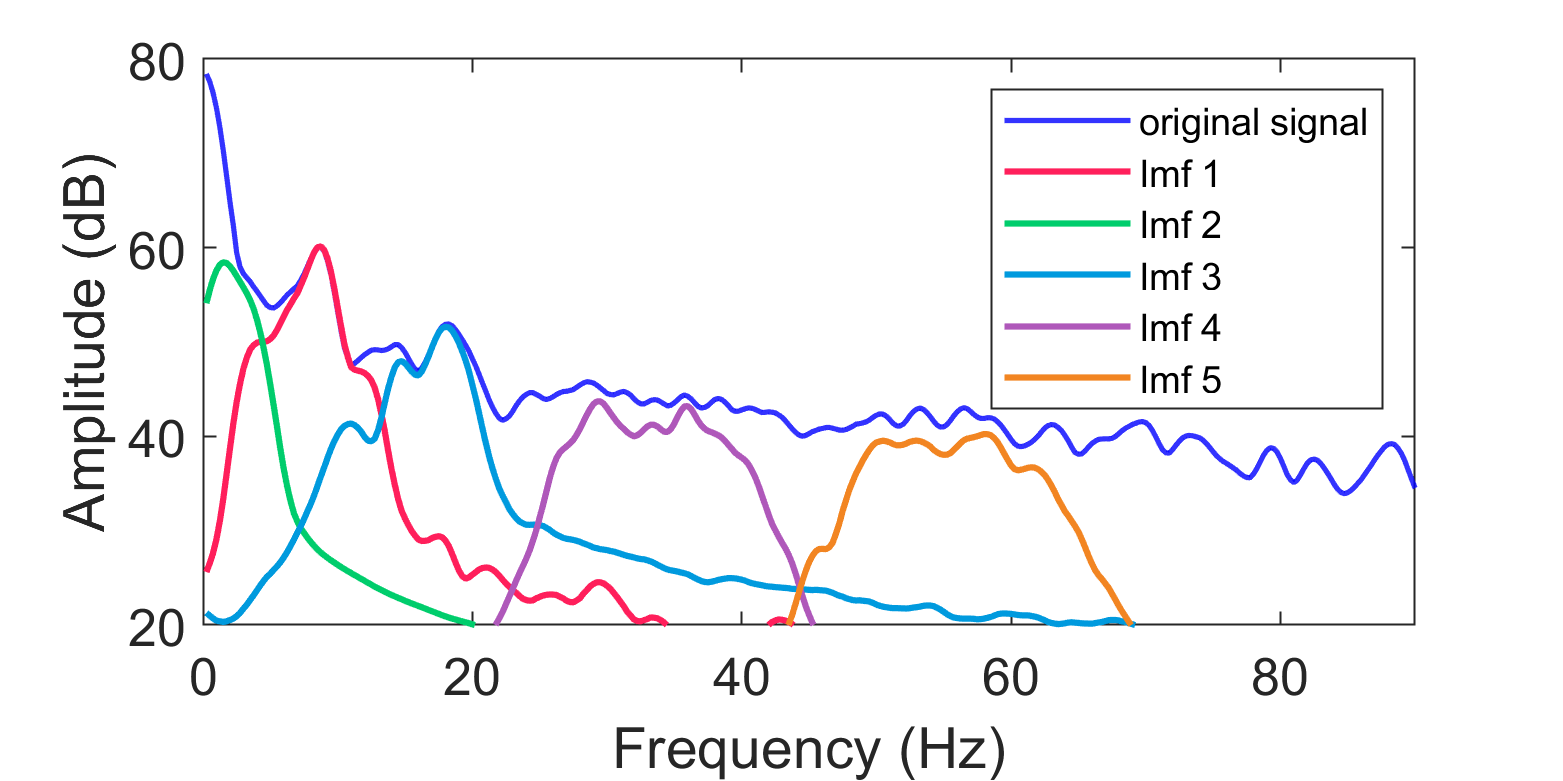} &   \hskip-1.0cm\includegraphics[width=.375\textwidth]{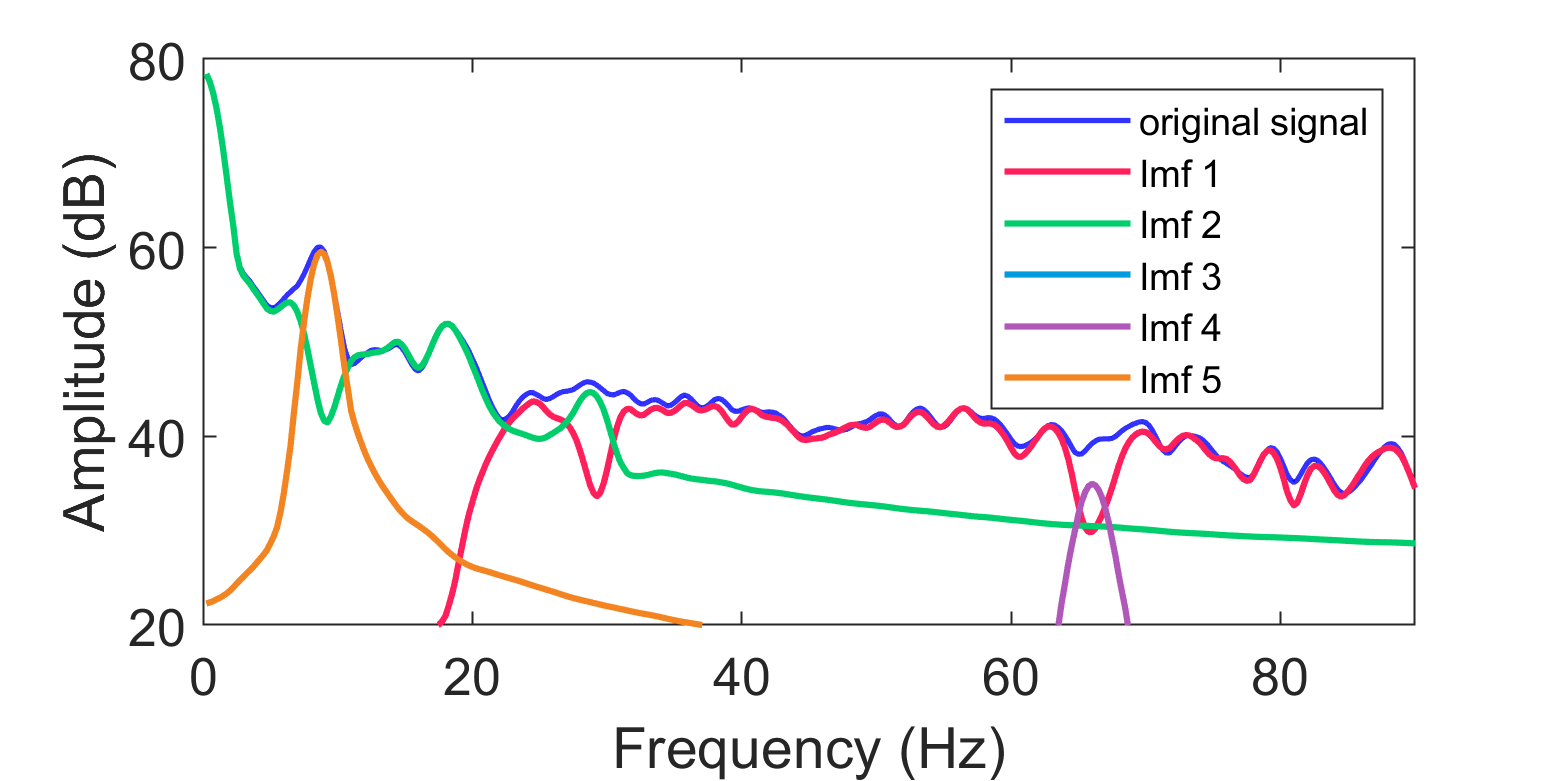} \\
(e) SST & \hskip-1.0cm (f) SSA \\
\multicolumn{2}{c}{ \includegraphics[width=.375\textwidth]{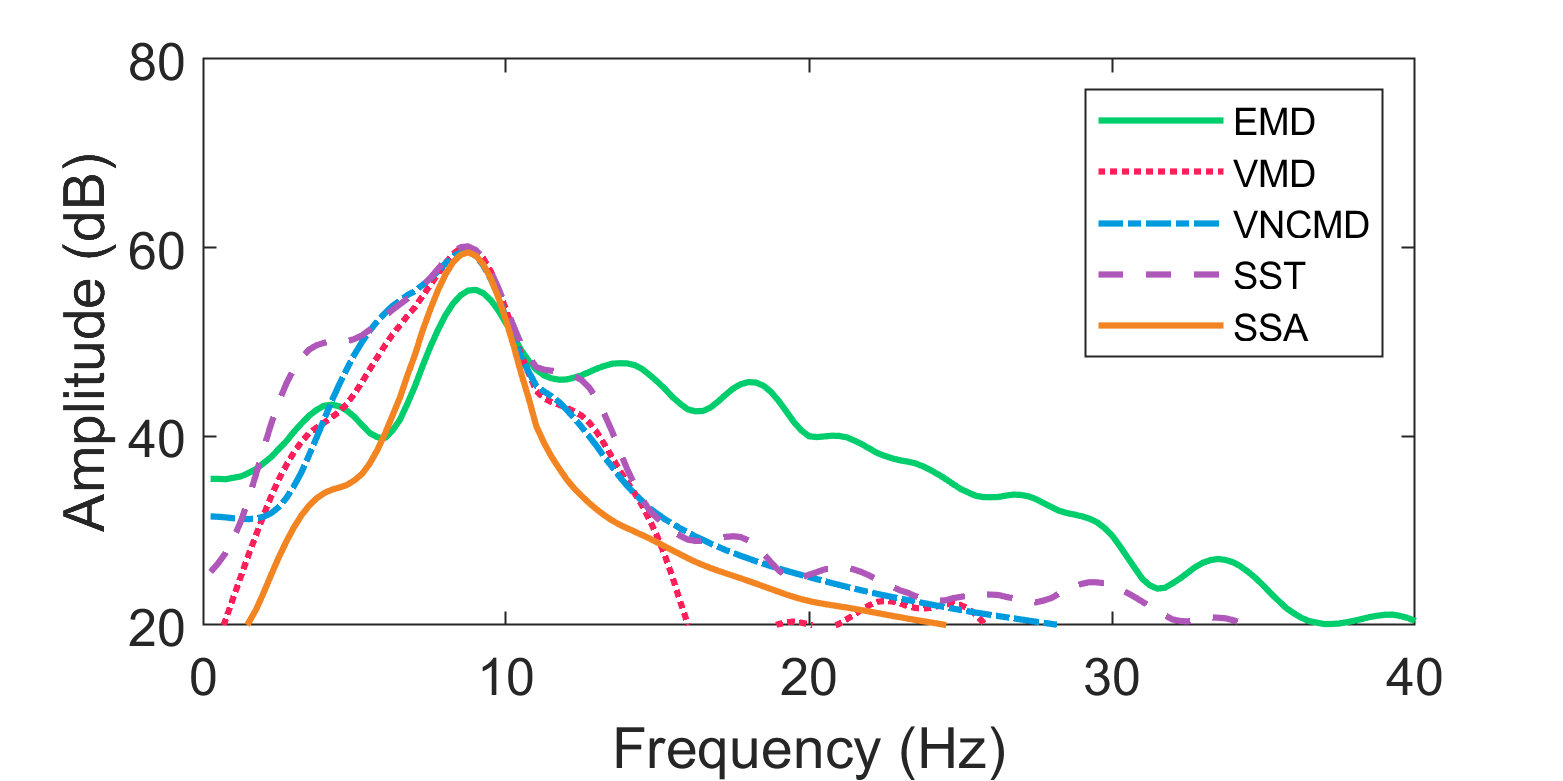}}\\

\multicolumn{2}{c}{ (g) Alpha frequency}
  \\
\end{tabular}
\caption{Results of SD methods applied on a real-life EEG data. The original time-series is shown in (a), while the spectra of the $K=5$ decomposed components are shown for EMD (b), VMD (c), VNCMD (d), SST (e) and SSA (f). The spectra of the relevant modes (corresponding to the $\alpha$ frequency range (8-10 Hz)) obtained from different methods are shown in (g).}
\label{figure:realworldsignal}
\end{figure}

This case study examines the performance of different SD methods on a real-life EEG data. EEG refers to a non-invasive technique that records brain's electrical activity over a period of time and is commonly used in clinical settings to diagnose epilepsy, sleep disorders and brain death. Here, we investigate the human EEG data recorded in a resting state - eyes closed. During the rest state, there is a pronounced EEG activity within the frequency range of 8-12 Hz - the so called alpha-rhythms. On the other hand, opening his/her eyes is marked by reduced alpha-rhythm in the EEG data. The data comprised of EEG recordings from a single subject who remained in the relaxed state with his eyes closed for four seconds, as seen in Figure \ref{figure:realworldsignal} (a). The data was recorded at COMSATS University Islamabad and obtained through an OpenBCI Cyton board at a sampling rate of 250 Hz \cite{Rehman19}.

Here, we applied different SD methods to the EEG data with an aim to extract the component corresponding to the alpha-rhythm. The smoothed spectra of the extracted components (named Imf1-Imf5 in the figure) obtained from EMD, VMD, VNCMD, SST and SSA methods are shown in Figure \ref{figure:realworldsignal} (b-f). In Figure \ref{figure:realworldsignal} (g), we show the plots of the extracted mode corresponding to the alpha-rhythm obtained from the different SD approaches.

The analysis of Figure \ref{figure:realworldsignal} suggests that all the methods were able to extract the component corresponding to the alpha-rhythm, albeit to the varying degree of accuracy. In the case of EMD, the relevant component, shown in orange in (b), demonstrates wide-band characteristics suggesting that the component contains artefacts in addition to the signal of interest i.e., alpha-rhythm. The spectra of the relevant components obtained from the VMD and the VNCMD methods, shown in green in (c) and (d) respectively, are relatively narrow-band signals peaking at the desired frequency range of 8-12 Hz - hence more accurate. In the case of SST, there appears to be significant overlap between the spectra of the adjacent components in addition to the spectrum of the relevant component being relatively wide-band. Both these observations suggest that the alpha-rhythm may not be extracted accurately by the SST. Finally, the best performance is delivered by the SSA method, as shown in (f). Not only are the extracted components separated from each other in the frequency domain but the relevant component (shown in orange) is narrow-band and ideally centered around the desired frequency range of 8-12 Hz. The superiority of the SSA method is further confirmed by examining the plot in (g) that shows the spectra of the alpha-rhythm obtained from different methods. It is clear that the SSA produced the alpha-rhythm component with the narrowest spectrum among all other methods.         

\section{Robustness to Noise}
\label{Section4}
Noise is ubiquitous in most real-life signals, masking the desirable information content in the data. It is therefore imperative to either remove noise from data as a preprocessing step or to design methods that are inherently robust to noise. SD and T-F approaches are routinely used in applications involving noisy data sets. To this end, we will investigate the performance of different data-driven SD methods in the presence of noise. 

\begin{figure}[!htb]
\centering
\centerline{\begin{tabular}{cc}
  \includegraphics[width=.47\textwidth]{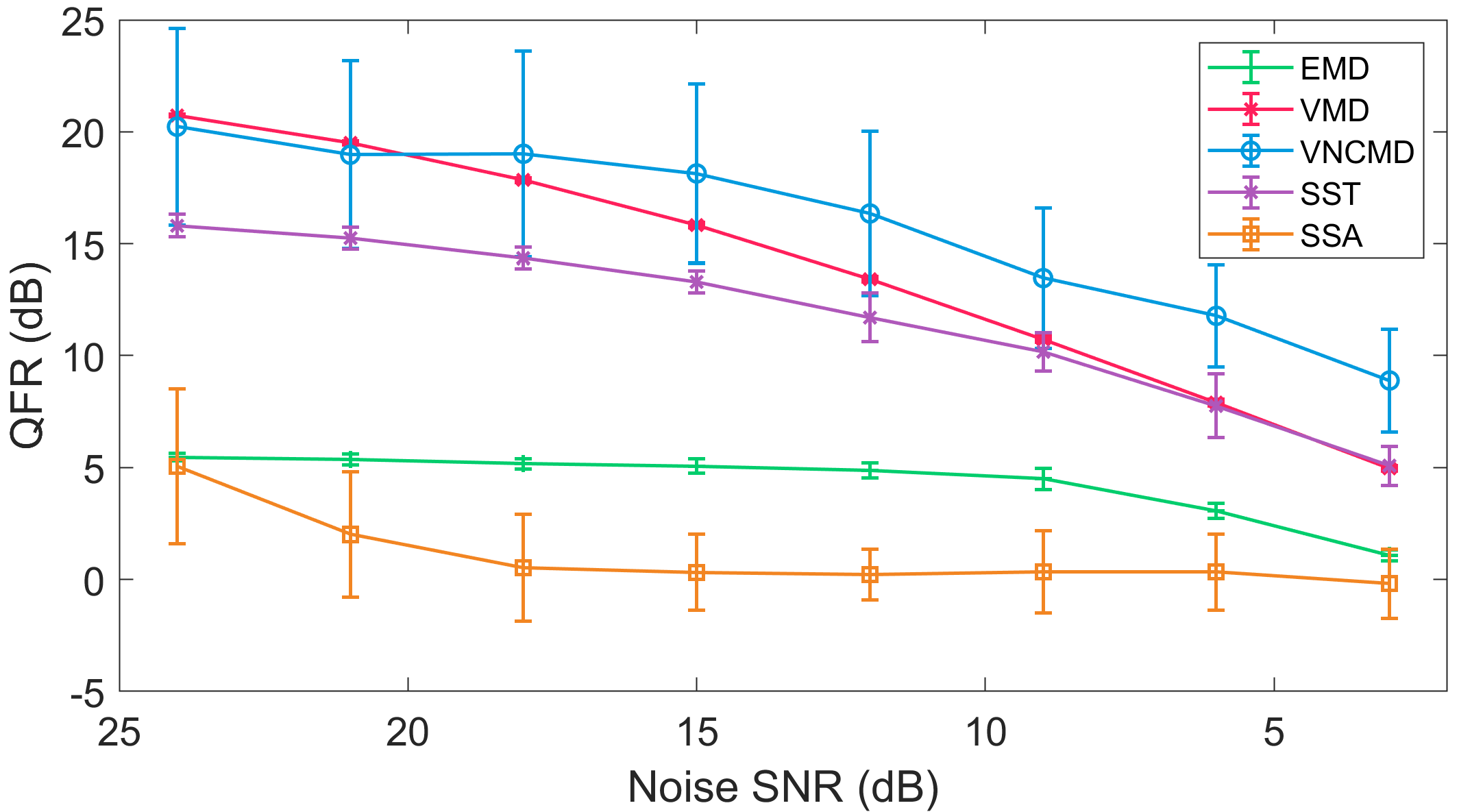} &   \hskip-0.0cm\includegraphics[width=.47\textwidth]{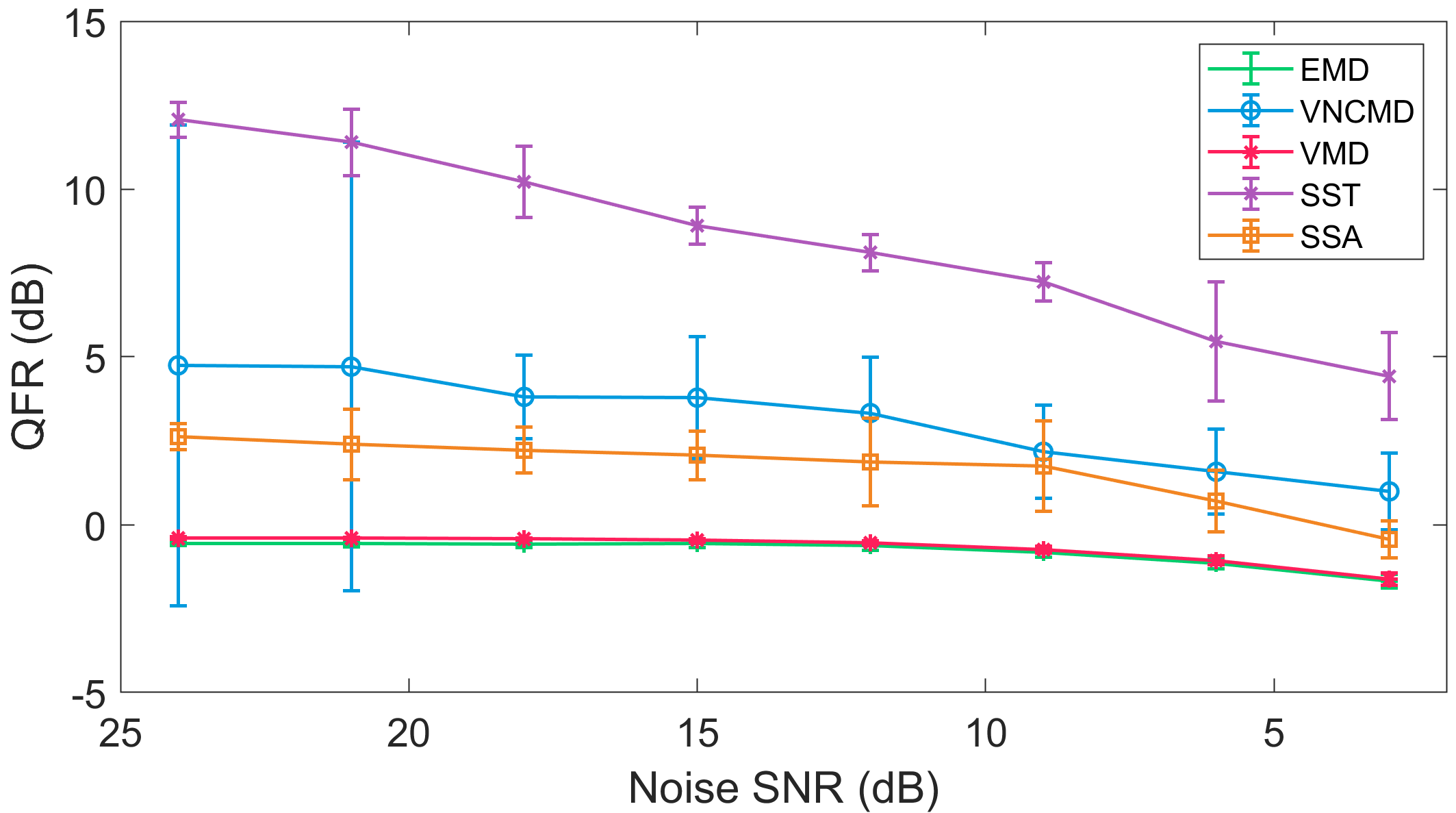} \\
(a) Synthetic signal 1 & \hspace{0.85cm} (b) Synthetic signal 2  \\
\end{tabular}}
\caption{QRF vs SNR for synthetic signal 1 (a) and synthetic signal 2 (b), averaged over 50 different iterations of Gaussian white noise. The standard deviation is plotted as error bars for each method}%
\label{fig:syntheticsignalsNoise}
\end{figure}

We employ the QRF measure to assess the accuracy of signal decomposition methods under noisy inputs. The noisy input is generated by adding the white Gaussian noise (wGn) with varying powers, corresponding to a range of SNR = 24 dB -- 3 dB, to $s_1$ (narrow-band signal) and $s_2$ (wide-band signal). We generated 50 realizations of wGn corresponding to each SNR value and obtained an ensemble of decomposed components from the relevant SD methods. Then, the QRF values were computed for those decomposed components (from multiple SD methods), for all 50 realizations of input noisy data. The mean QRF value along with its standard deviation is computed across all 50 realizations for each decomposed component obtained from multiple SD methods. Finally, the QRF values of all decomposed components from a particular SD method were summed to obtain a single QRF value corresponding to each method. We plot those values in Figure \ref{fig:syntheticsignalsNoise} (a) and (b) for $s_1$ and $s_2$ respectively, against a range of input SNR.

For the noisy $s_1$, VNCMD performed the best for the lower range of SNR values but exhibited large deviation in its performance as compared to all other techniques. The means that the performance of the VNCMD was erratic: performing very well on some noisy signals but  significantly worse on others. VMD was found to be very robust to noise and demonstrated the lowest performance deviation across different noise realizations. SST also performed quite well across all the input SNR values. Note that the discontinuity in the T-F plane of $s_1$ contributed to the low QRF values for SST as the method inherently is not designed to obtain components corresponding to the discontinuous T-F ridges. By removing the discontinuity, we found that the SST performed on par with the VMD. The SST's performance was also very stable across different noise realizations as depicted by its low standard deviation range. EMD and SSA performed considerably worse than the other methods as highlighted by the very low QRF values. This was somewhat expected as both these techniques also did not perform well in the absence of noise (see Figure 2 and 3). 

For the wide-band signal $s_2$, the SST was the best performing method, not only in terms of higher QRF values but also the lower standard deviation of the QRF across all nose realizations. The VNCMD was again very unpredictable showing very large performance deviations especially at high SNR values. This is in addition to the fact that the VNCMD requires initial component frequencies to be set very close to the `ground truth' for any meaningful results. Not surprisingly, the performance of the EMD, VMD and SSA was poor owing to the wide-band nature of the input signal $s_2$. With EMD and VMD, however, the performance variation across multiple noise realizations was minimal.

Overall, the SST was found to be the most robust technique against noisy data. This was expected since SST is the only SD approach that comes with theoretical guarantees regarding its robustness against noise \cite{Thakur13}.       

\section{Sensitivity to Changes in  Parameters}
\label{parameterSensitivity}
This section describes the dependence of the SD methods on their algorithmic parameters and how sensitive or robust they are to the changes in parameters.

The \textbf{VMD method} is dependent on the following parameters: i) the number of components $K$ to be extracted; ii) $\alpha$ that dictates the bandwidth of the extracted components; iii) $\tau$ that determines the signal reconstruction accuracy. 

Choosing the right number of components $K$ is important for the operation of VMD. Not doing so typically results in either mode-mixing or mode-splitting in the decomposed components. Several recent works address this issue: for instance, successive VMD \cite{Nazari19} obtains the components in an iterative fashion and therefore does not require the user-defined parameter $K$. Further, adaptive extensions of the VMD method \cite{Qiming21} have also recently surfaced.   

The $\alpha$ parameter also influences the result of the signal decomposition, but the range of its viable values is typically quite large. For instance, in our experiments, no discernible difference in the VMD output was found for the test signals $s_1$ and $s_2$, for the range of $\alpha=50-1000$. For the values above $\alpha=2000$, the decomposition accuracy began to suffer for the two test signals. For the EEG signal, the values of $\alpha=2000$ produced the best output as the narrow-band alpha-rhythm was the desired signal. It is noteworthy that the optimal value for $\alpha$ will depend on the input signal and application at hand. Higher values of $\alpha$ reduce the bandwidth of the extracted components and may increase the risk of not capturing the correct center frequencies. On the other hand, lower values of $\alpha$ tend to increase the bandwidth of the components and may introduce artefacts and mode-mixing in the output. 

Lastly, setting $\tau = 0$ is optimal for signals with noise as complete reconstruction is not needed in this case. For the non-noisy signals, setting $\tau$ larger than zero but smaller than 1 yielded more accurate results.

Being an extension of the VMD method for wide-band signals, \textbf{VNCMD} shares some of the parameters of the VMD method e.g., the number of components $K$ and the bandwidth related parameter. However, the method introduces several new user-defined parameters/initial conditions, proper tuning of which is critical to the  operation of VNCMD. For example, the specification of the initial values for the center frequencies of the extracted components is an important parameter. In our experiments, for $s_1$, the initial values were chosen to be 85 Hz, 50 Hz and 30Hz, for component 3, 2 and 1 respectively; but changing any of those by just 5 Hz significantly deteriorated the results. As an example, changing the initial frequency of component 3 from 85 Hz to 90 Hz resulted in the corresponding QRF dropping from 24.4 to 4.4 dB.

Unlike VMD, the VNCMD was found sensitive to variations in the $\alpha$ parameter that dictates the bandwidth. Further, the method needs tuning of another parameter $\mu$ that relates to the update of the instananeous frequency (IF) within VNCMD. In our experiments, we found the VNCMD method to be very sensitive to changes in $\mu$ so much so that in some cases the method even failed to converge. Overall, the VNCMD method suffers from serious convergence issues and even in cases where it converges, it is hard to tune the algorithmic parameters to yield optimal results.

In \textbf{SST}, an important parameter is the number $K$ of decomposed components. Further, since SST involves taking the wavelet (or short-time Fourier) transform of the signal, specifying related parameters, e.g., the mother wavelet (e.g., Morlet), the number of scales (frequencies), and $\mu$ that dictates the time-frequency trade off, is an important step. Following the application of the time-scale (frequency) transform and squeezing of the resulting spectrum, the next step within SST is the curve extraction which affects the signal decomposition results via SST. The important parameters governing the curve extraction process include \textit{StartBand} which set the initial width of the band around each center frequency and \textit{MaxStepSize} which restricts the maximum amount of change in center frequency as the curve extraction goes forward in time. In our experiments, we obtained good results by choosing the above two parameters within the range of 5-30. For noise robustness, the SST utilizes a threshold parameter $\gamma$; in our experiments, in our experiments involving noise robustness, we obtained optimal results by setting $\gamma = 10^{-6}$.

The \textbf{SSA} was found to be very sensitive to parameter changes. Like VMD and SST, SSA also requires users to specify the number of decomposed components a priori. The embedding dimension $L$ is another parameter that determines how well the components are separated. For the test signals used in our study, even small changes in $L$ produced results of varying accuracy. We found that the range of $L=30-150$ produced decent results, with higher values increasing the computation time but not necessarily the accuracy (QRF). The threshold parameter $\epsilon$ that affects robustness against noise was set in the range of $\epsilon = 10^{-5}$ to $10^{-7}$ to provide good results.

The \textbf{EMD} method is advantageous in a sense that it does not require a priori specification of the number of decomposed components. An considerations with the EMD method is the choice of interpolation function for signal extrema. For that purpose, cubic spline interpolation is by far the most widely used technique within EMD. Another important factor in EMD operation is the stopping criterion for decomposed components (also known as intrinsic mode function or IMF). We use a popular criterion, introduced in \cite{EMD-alg}, which uses two threshold values to stop the sifting process within EMD.     

\section{Multivariate Signal Decomposition}
\label{multivariate}

Multivariate signals comprise of multiple data channels (time-series) that may be correlated to each other. While processing such signals, it is vital to consider inherent correlation among multiple data channels. That requires representing and viewing a multivariate signal in multidimensional space where it naturally resides. With this representation, the decomposition of multivariate signals amounts to the extraction of inherent \textit{rotational} components (instead of oscillatory components for univariate signals) in multidimensional spaces. This principle forms the crux of two of the most popular multivariate signal decomposition approaches, multivariate EMD (MEMD) \cite{Rehman10} and multivariate VMD (MVMD) \cite{Rehman19}, which will be evaluated in this section. The details of the operation of the two algorithms as well as the illustration of the multivariate rotational components obtained from those can be found in the relevant articles \cite{Rehman10, Rehman19}.

In particular, we shall investigate and evaluate the mode-alignment property of MEMD and MVMD in this section. We shall also examine how the addition of noise in multivariate data affects the mode-alignment property. The mode-alignment refers to the matching of modes with similar frequency content across multiple channels in the same indexed components. This property is considered a prerequisite for a wide range of engineering applications involving non-stationary multivariate signals, e.g., data fusion \cite{Rehman09-fus}, denoising \cite{Naveed20} and biomedical signal classification \cite{Park13}.      

\subsection{Mode-alignment}
This section will explore the mode-alignment properties of the two popular multivariate SD methods, MEMD and MVMD, and compare those against a univariate method (VMD) to highlight the superiority of the multivariate approaches. We will consider both synthetic and real-life biomedical signals to support our analysis. 

\subsubsection{Case Study 1: Multivariate Synthetic Signal}

The first case study involved a synthetic bivariate (two-channel) signal containing a combination of two components in the first channel, a 2 Hz and a 50 Hz sinusoid; and a combination of three oscillatory components comprising of a 2 Hz, 20 Hz and a 50 Hz sinusoid in the second channel. The correct mode-alignment of the decomposed data should result in three decomposed components with the first component containing the low-frequency 2 Hz component in both channels. The second decomposed component should contain a 20 Hz sinusoid in channel two with no signal in channel 1. Similarly, the third decomposed component should contain a 50 Hz sinusoid in both the channels. 

\begin{figure}[!htb]

    \centering
    \centerline{\begin{tabular}{ccc}
      \includegraphics[width=.33\textwidth]{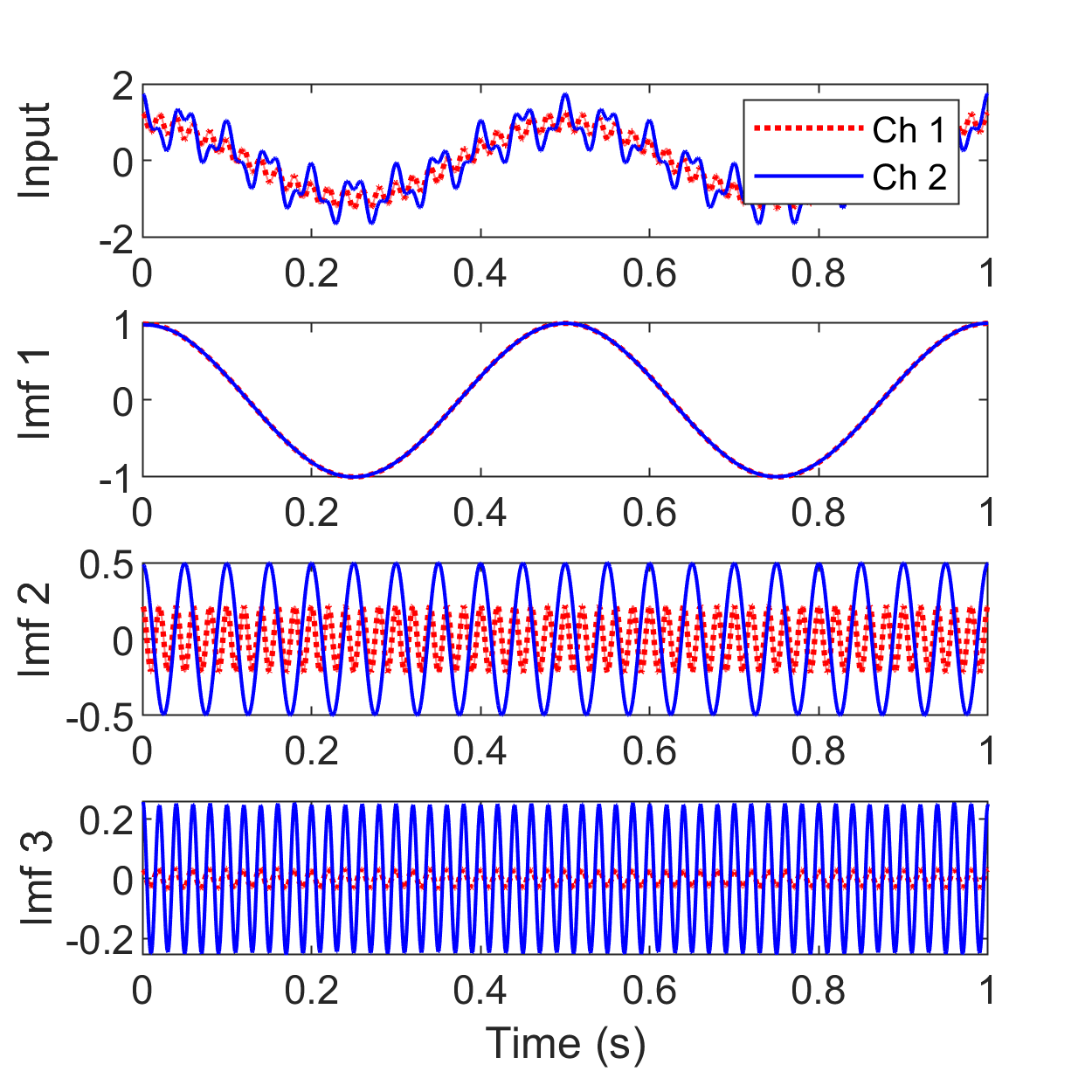} & 
      \hskip-0.7cm\includegraphics[width=.33\textwidth]{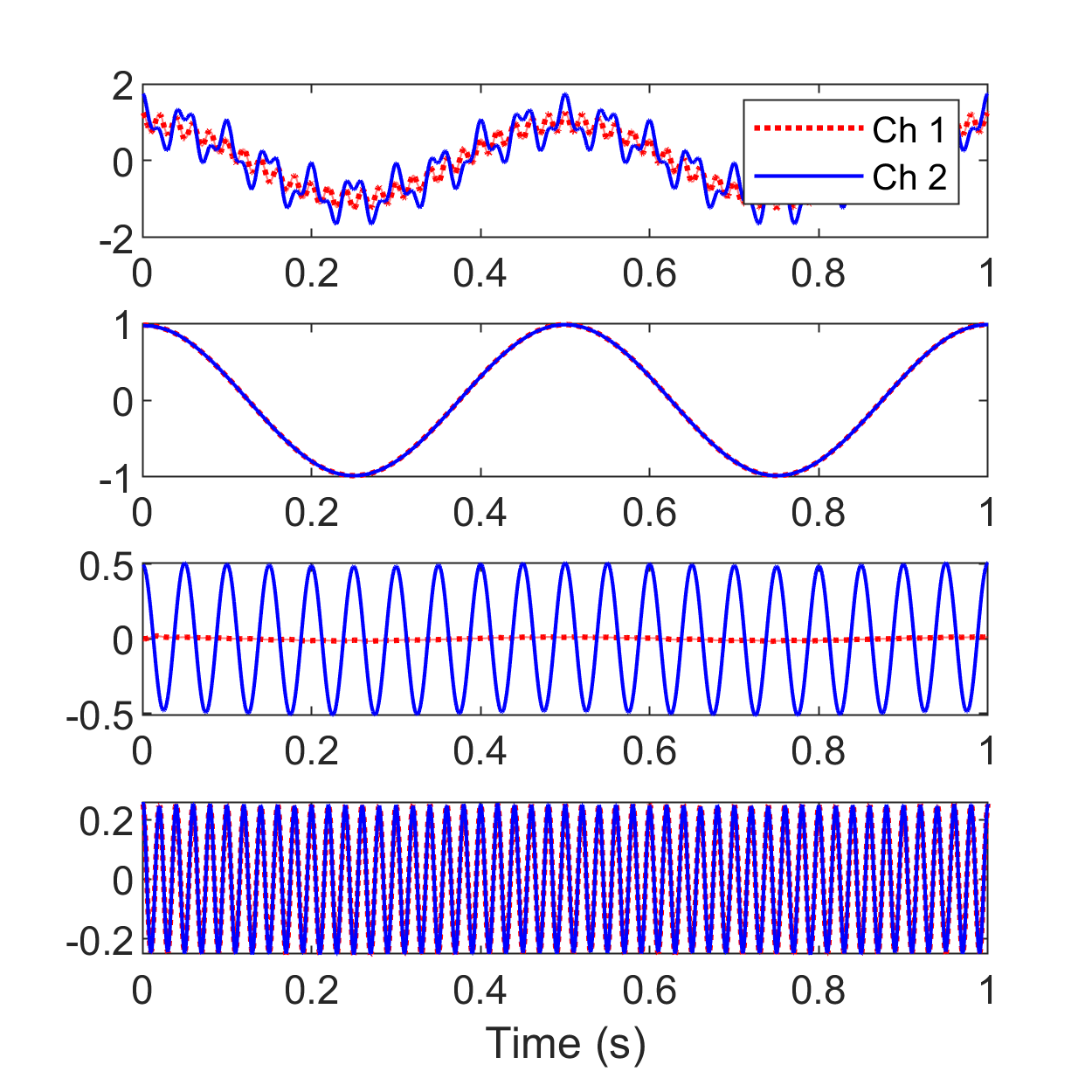} & \hskip-0.7cm\includegraphics[width=.33\textwidth]{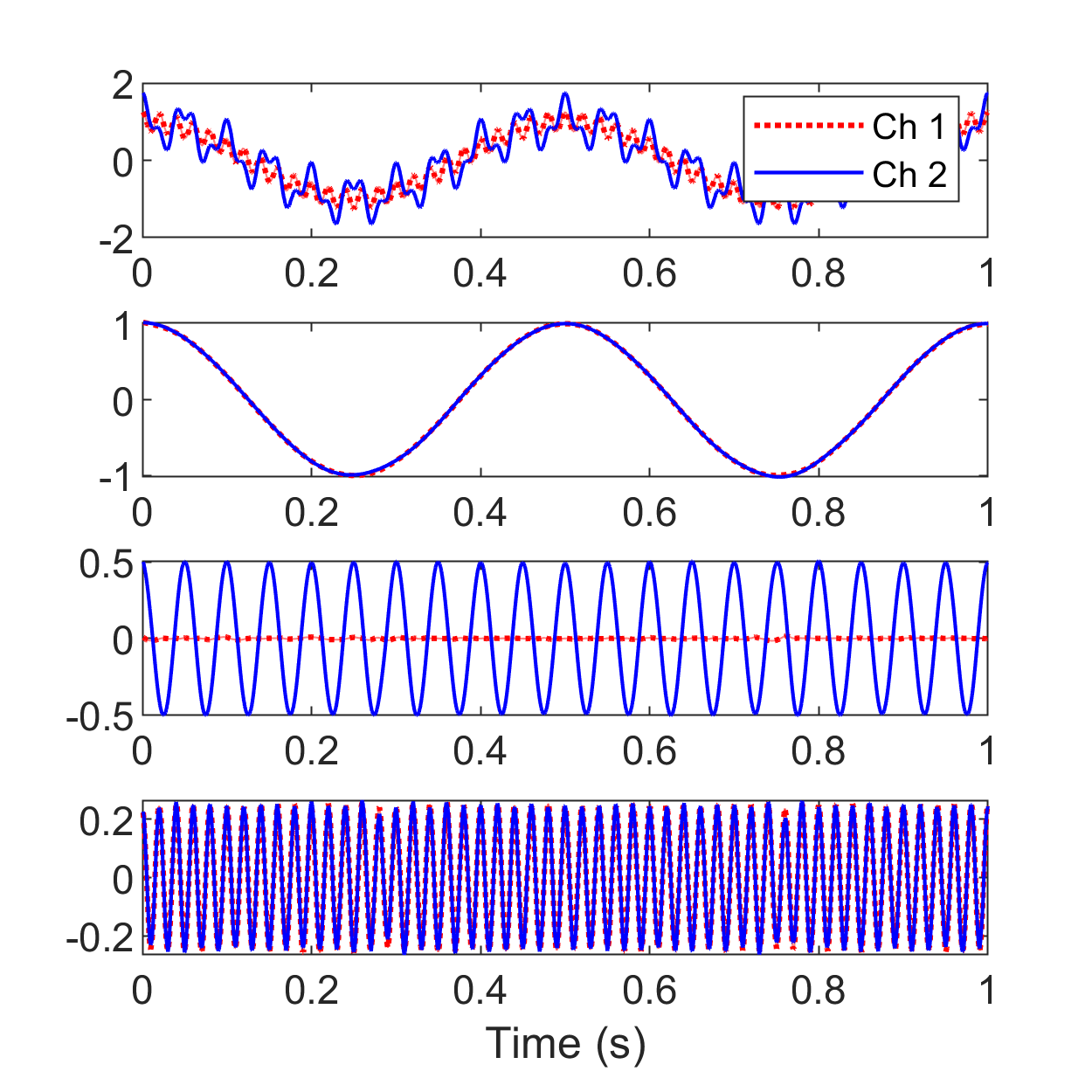} \\
      \hskip-0.7cm\small{\textbf{(a)} VMD with 40dB SNR}  & 
      \hskip-0.7cm\small{\textbf{(b)}  MVMD with 40dB} & 
      \hskip-0.7cm\small{\textbf{(c)} MEMD with 40dB}  \\
       
      \includegraphics[width=.33\textwidth]{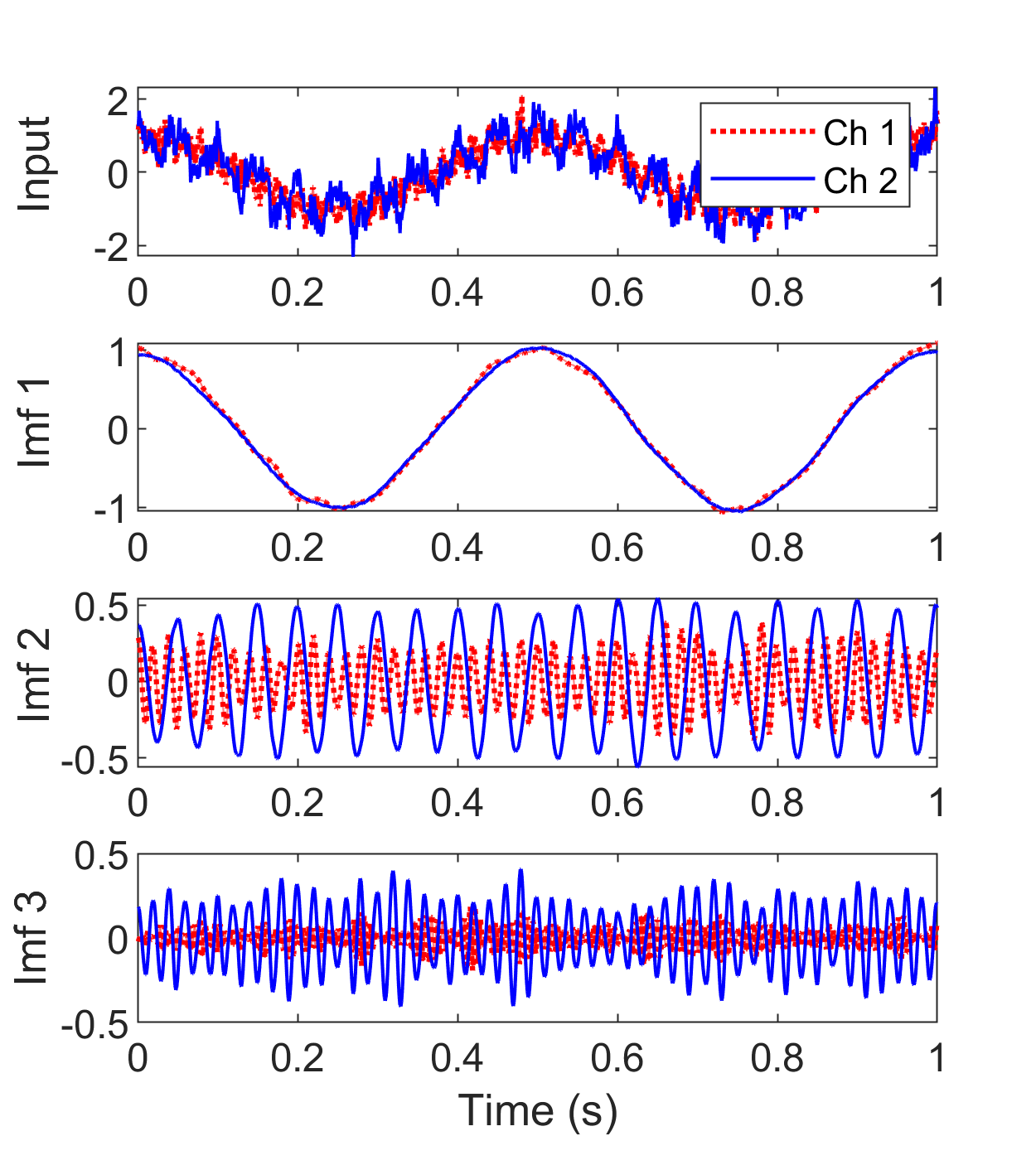} & 
      \hskip-0.7cm\includegraphics[width=.33\textwidth]{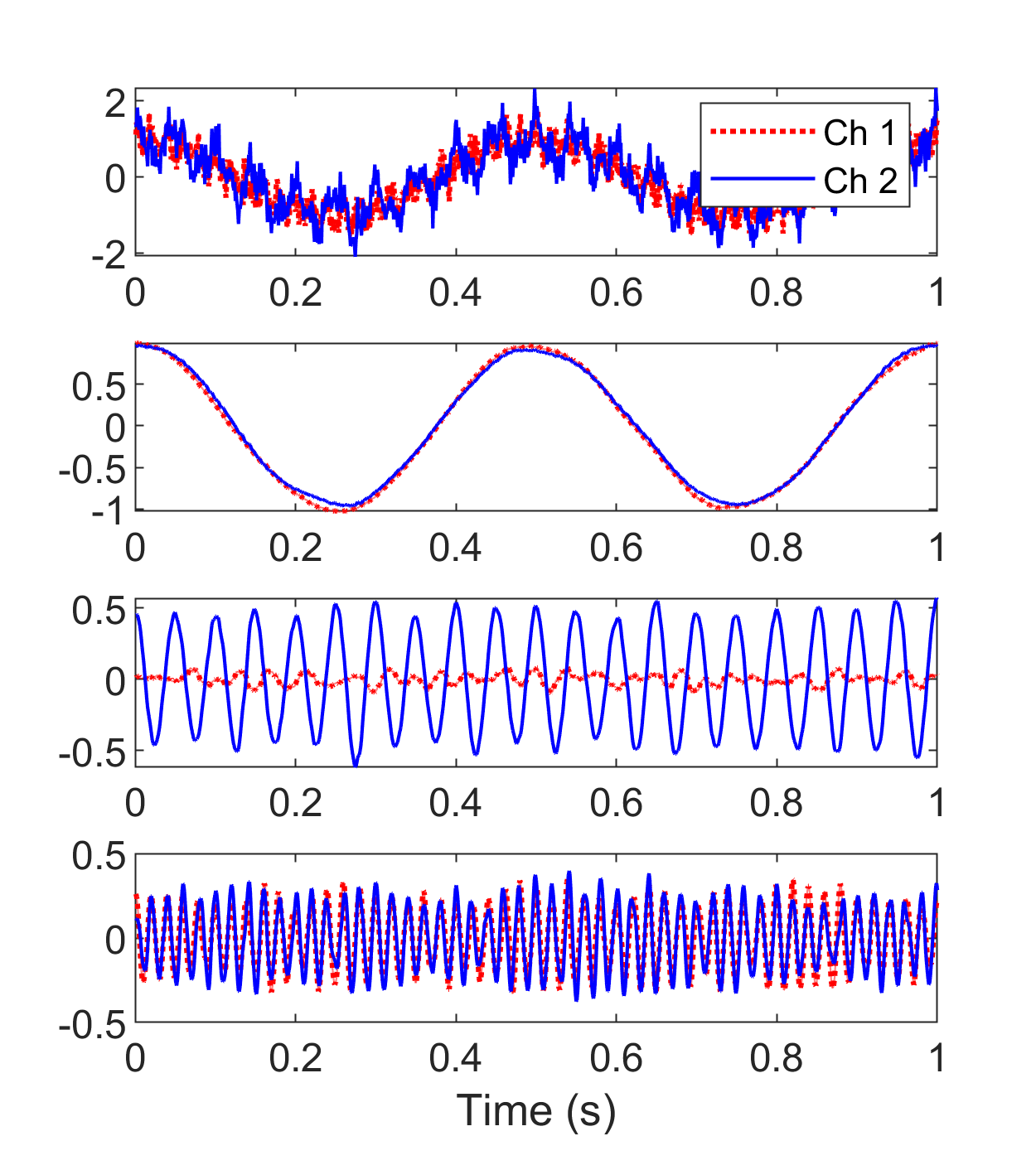} & \hskip-0.7cm\includegraphics[width=.33\textwidth]{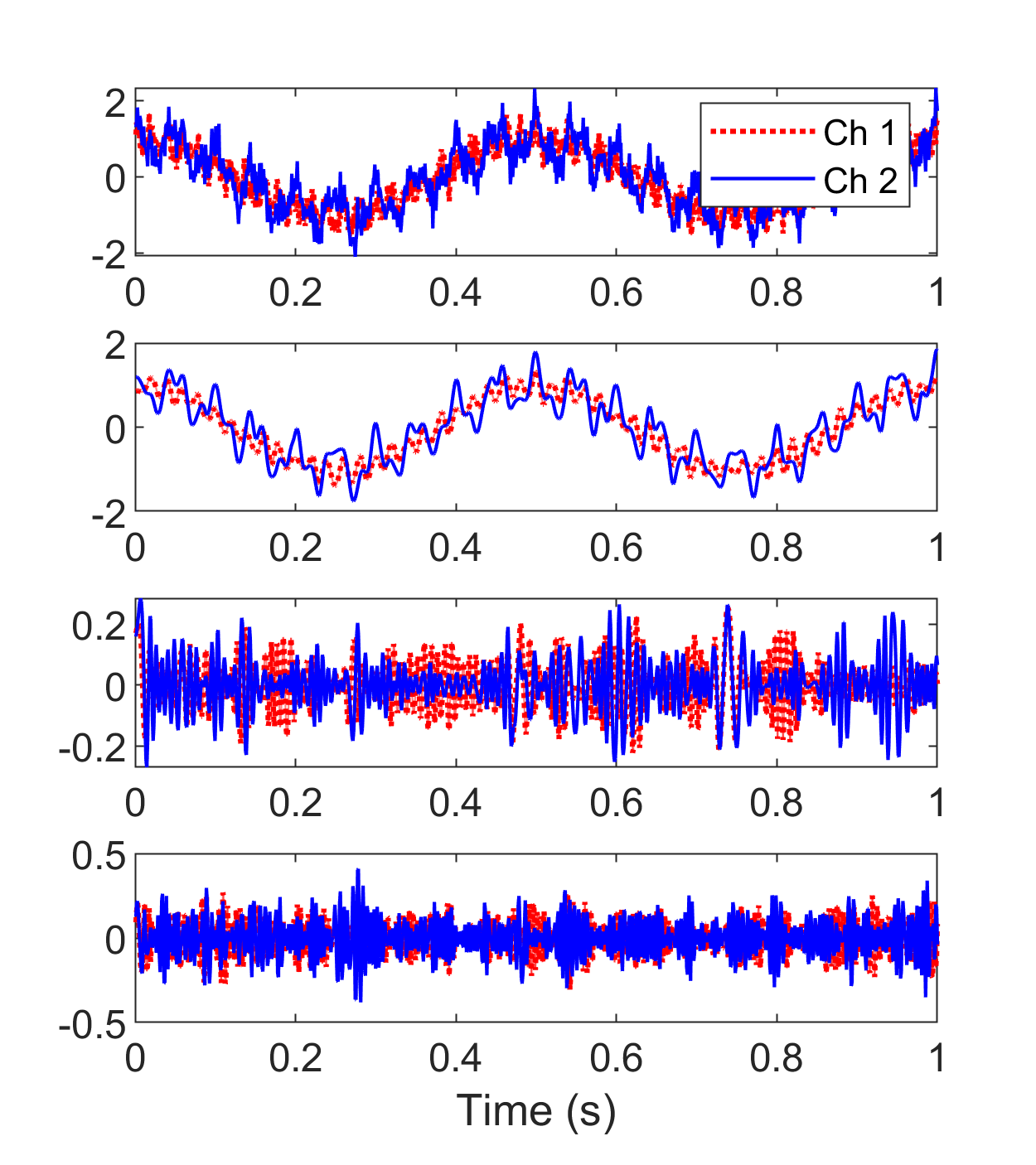} \\
      \hskip-0.7cm\small{\textbf{(d)} VMD with 10dB SNR}  &
      \hskip-0.7cm\small{\textbf{(e)}  MVMD with 10dB SNR} &
      \hskip-0.7cm\small{\textbf{(f)} MEMD with 10dB SNR}  \\
    \end{tabular}}
    
    \caption{Signal decomposition of noisy multivariate synthetic signals, corresponding to the $SNR=40dB$ (top row) and $SNR=10dB$ (bottom row) with white Gaussian noise, obtained by applying the VMD (channel-wise), MVMD and MEMD algorithms.}
    \label{fig:modeallignment_synthetic}%
\end{figure}

The Figure \ref{fig:modeallignment_synthetic} (a-c) show the signal decomposition results of the noisy multivariate synthetic signal corresponding to the $SNR=40dB$ obtained by applying the VMD (a), MVMD (b) and MEMD (c) methods. It is clear that the VMD could not properly align the sinusoids across different channels in the components 2 and 3: specifically, the decomposed component in mode 2 contains two signals with different frequencies, i.e. 20 Hz and 50 Hz. On the other hand, the MVMD and the MEMD methods were both successful in decomposing the three modes with correct sinusoids, showcasing mode-alignment capability. 

We next increased the input noise strength in all channels of the multivariate signal corresponding to the $SNR=10dB$ and applied the three methods again to obtain the signal decomposition. The results are shown in Figure \ref{fig:modeallignment_synthetic} (d-f) for VMD, MVMD and MEMD respectively. It is apparent that at higher input noise levels, the performance of the MEMD deteriorates significantly with no mode-alignment present in any of the three decomposed components. It was observed (while not shown here) that the MEMD failed to decompose multivariate signals properly even at the SNR level close to 30dB. Contrarily, MVMD was found to be robust to noise, delivering impressive results in terms of mode-alignment at the $SNR=10dB$ as shown in Figure \ref{fig:modeallignment_synthetic} (e). The decomposition results obtained by applying the VMD algorithm separately on multiple channels of the noisy multivariate synthetic signal are also shown in Figure \ref{fig:modeallignment_synthetic} (d). It can be noticed that the mode-alignment is absent in this case owing to the fact that the VMD does not operate in multidimensional space where the signal resides. 

\subsubsection{Case Study 2: Cardiotocographic data}
\begin{figure}[!htb]
    \centering
    \includegraphics[width=0.55\textwidth]{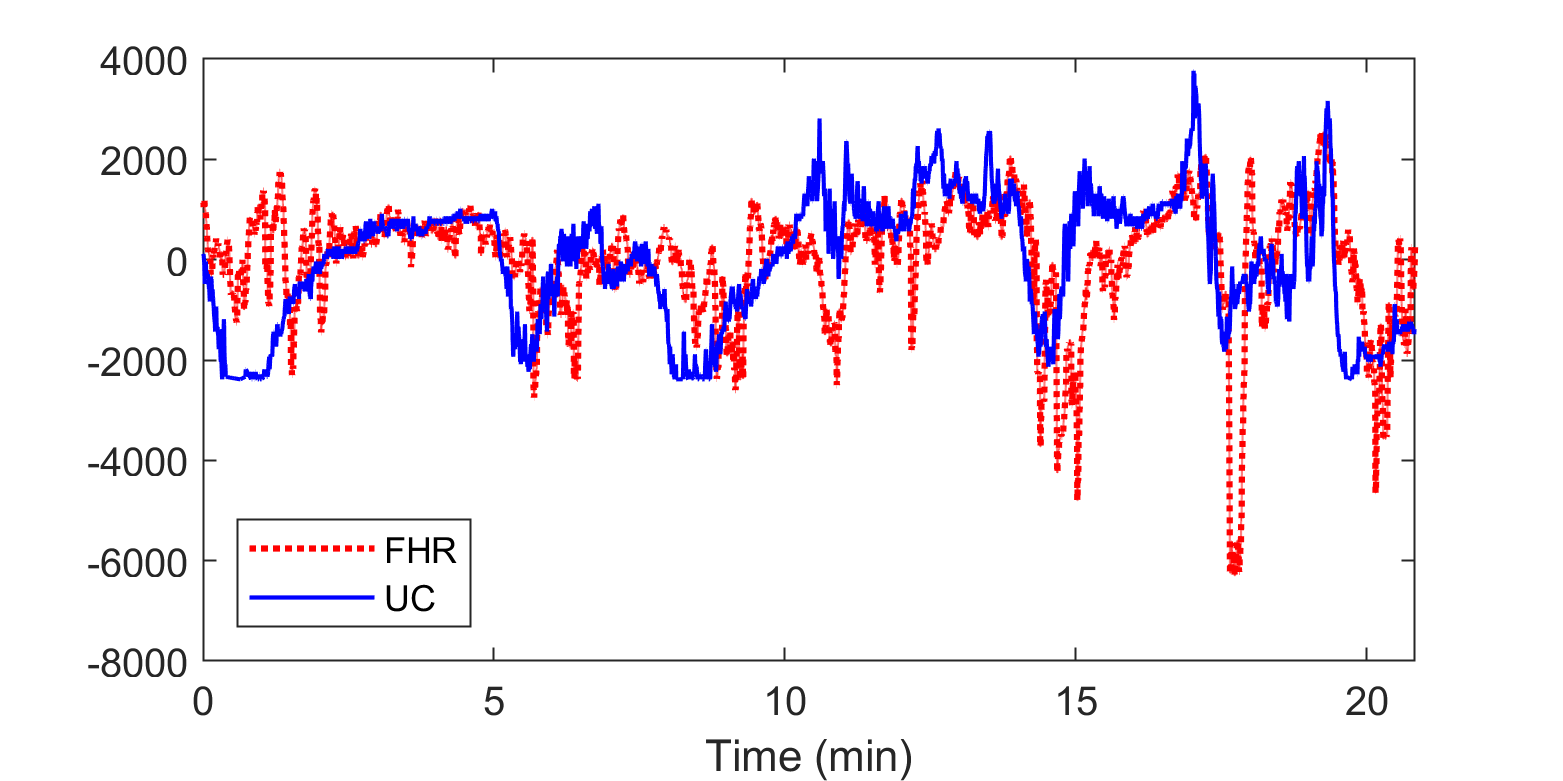}%

\caption{Time plot of cardiotocographic (CTG) recording of Fetal Heart Rate (FHR) and Maternal Uterine Contraction (UC).}%
    \label{fig:CTG}%
\end{figure}

The second case study for the qualitative evaluation of multivariate SD approaches in terms of mode-alignment involved a bivariate Cardiotocographic (CTG) data of Fetal Heart Rate (FHR) and Maternal Uterine Contraction (UC), as shown in Figure \ref{fig:CTG}. The recordings were taken from the freely available CTU-UHB intrapartum cardiotocography database by Physionet \cite{Physionet}. 

\begin{figure}[!htb]
    \centering
    \centerline{\begin{tabular}{ccc}
      \includegraphics[width=.35\textwidth]{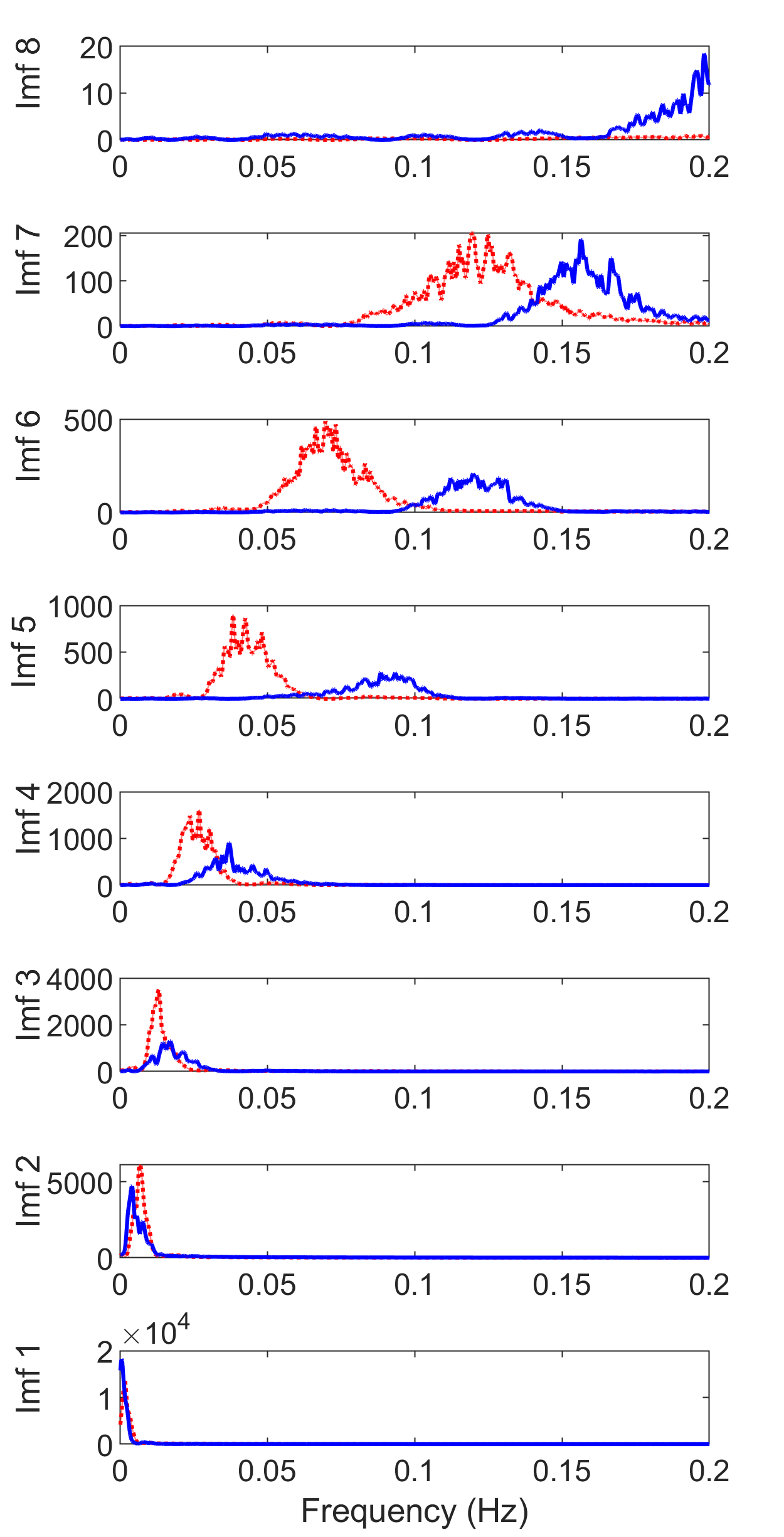} & 
      \hskip-0.7cm\includegraphics[width=.35\textwidth]{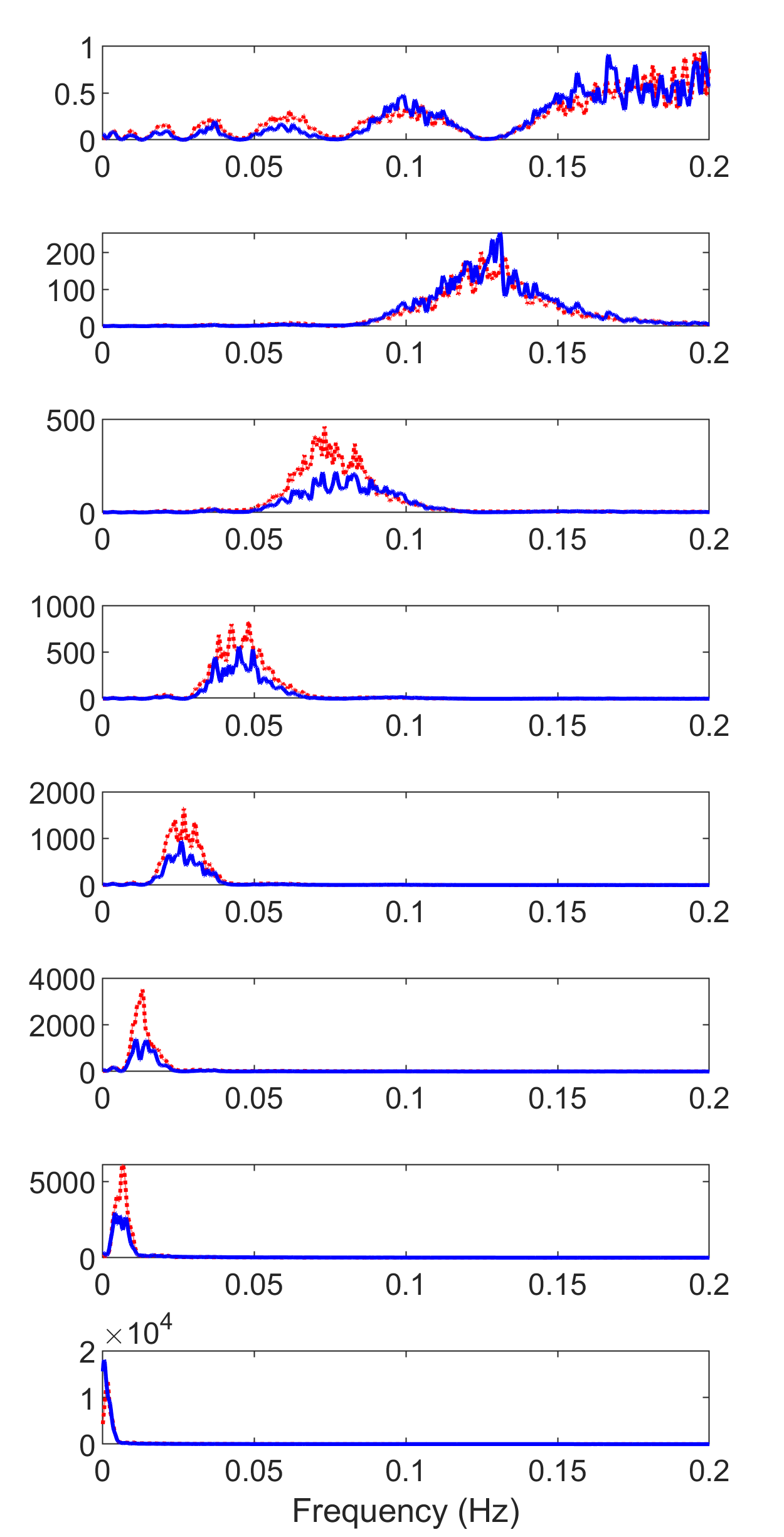} & \hskip-0.7cm\includegraphics[width=.35\textwidth]{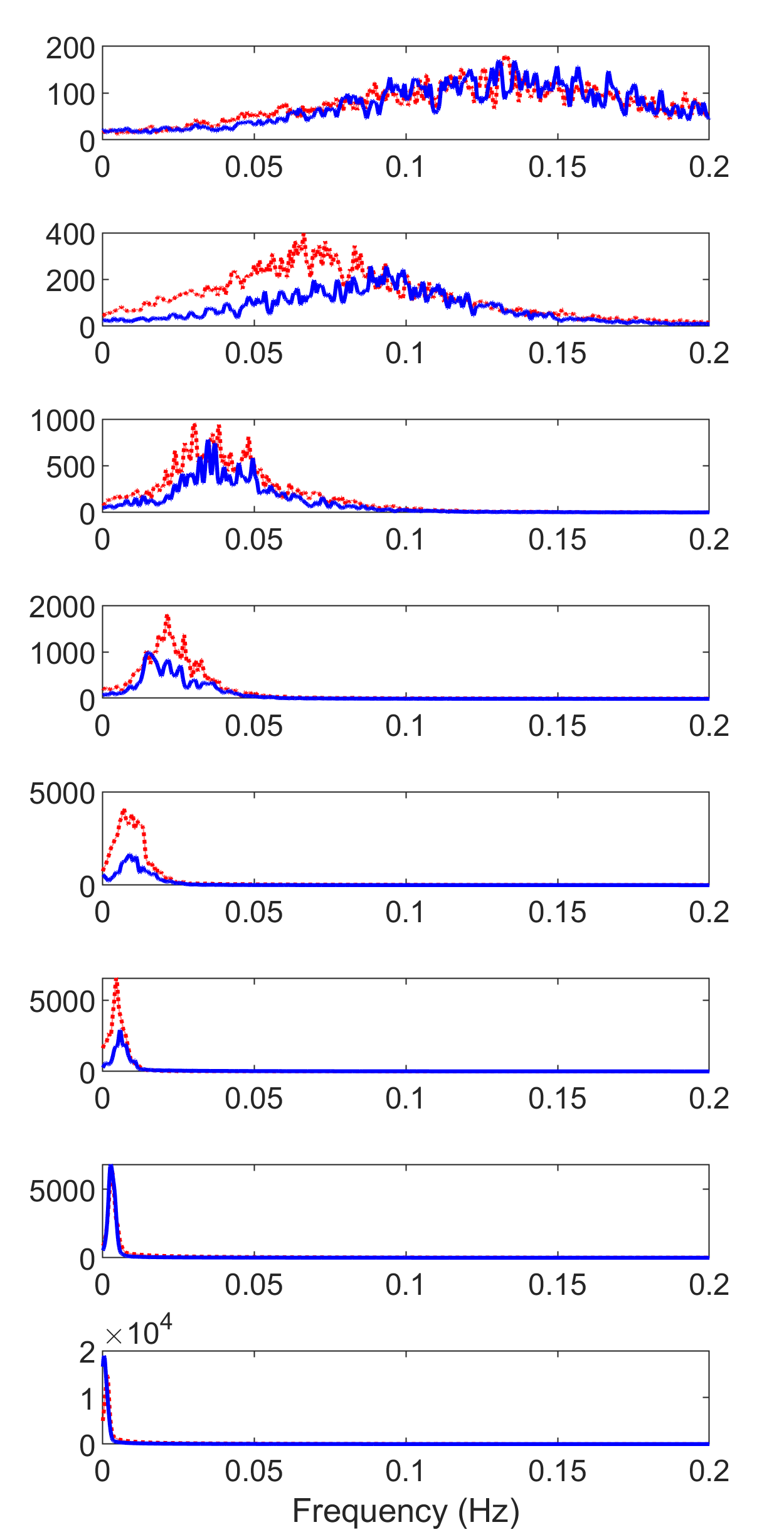} \\
    \small{\textbf{(a)} VMD} & 
    \hskip-1.0cm \small{\textbf{(b)} MVMD} & 
    \hskip-1.0cm \small{\textbf{(c)} MEMD}  \\
    \end{tabular}}
    
    \caption{Demonstration of mode-alignment capability of multivariate SD approaches on real-life biomedical data. The mode-alignment is not present in the case of VMD (a) whereas the decomposition obtained via MVMD (b) and MEMD (c) shows proper mode-alignment, with MVMD showing better mode-separation than MEMD.}%
    \label{fig:modeallignment_realCTG}%
\end{figure}

Figure \ref{fig:modeallignment_realCTG} shows the Fourier spectrum of the decomposed modes, extracted using the VMD (a), MVMD (b) and MEMD (c) methods. We set the total number of modes retrieved components to be equal to $K=8$. It can be noticed that the decomposed components from the VMD method (shown in the first column) are mostly free of artifacts related to mode-mixing as each mode consists of a narrow-band signal. That said, VMD components show poor mode-alignment across channels in almost all the decomposed components, as exhibited by misalignment of the spectra of the two channels in each mode.

The decomposed components obtained from MVMD, shown in Figure \ref{fig:modeallignment_realCTG} (b), exhibit accurate mode-alignment together with the narrow band signals -- indicating minimal mode-mixing. Another highlight of the MVMD is that the multiple decomposed components are non-overlapping in the frequency domain. This is not the case with MEMD which obtains decomposed components having relatively large bandwidths with overlapping spectra. However, the mode-alignment across multiple channels is also apparent in the case of MEMD with the exception of mode 7 where spectra of the two channels are not well-aligned.           

\subsection{Sensitivity to Changes in Parameters.}
\label{parameterSensitivityMultivariate}
Like EMD, the most important parameters affecting the performance of MEMD relate to the choice of the scheme used for extrema interpolation. We employed the widely used cubic spline interpolation in our experiments. For the decomposition of multivariate signals, MEMD takes multiple uniform projections of input signal in multidimensional space. The number of those projections is a user-defined parameter within MEMD. Fortunately, we found that the MEMD was rather insensitive to the changes in that parameter: we used $M=64$ projections in our experiments but found that any value of $M\geq 8$ yielded approximately similar results.   

The MVMD uses the same parameters as VMD, making the discussion in section \ref{parameterSensitivity} also relevant here. That said, in MVMD, we need to initialize the center frequencies $\omega_c$ of all channels. In our experiments, we found that initializing those to zero worked well in all the experiments that were conducted in this study.

\begin{table}
	\includegraphics[width=1\linewidth]{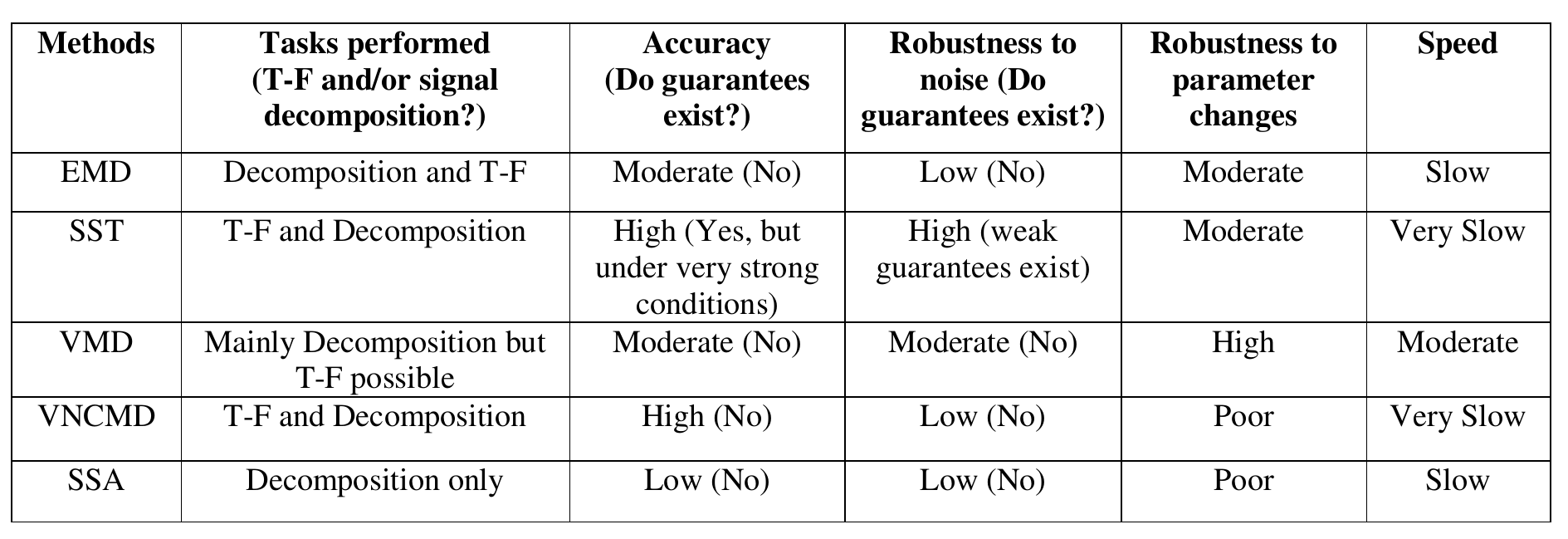}
\centering
	\caption{Comparing the performance of the data-driven SD approaches in terms of task performed, accuracy, robustness to noise, robustness to changes in parameters and practical speed.}	
\vspace{-5mm}
\end{table}

\section{Discussion}
In this section, we comment on the overall strengths and weaknesses of the evaluated SD methods in light of the insights developed through the experiments conducted in this study. In particular, we will evaluate their performance in terms of accuracy in signal (mode) decomposition, robustness to noise, sensitivity to parameter changes and their computational complexity. A summary is provided in Table 1, in light of the our experimental  observations regarding the performance of the most popular SD approaches. The table is self-explanatory, however, few observations on the major drawbacks of the existing approaches are in order.

In terms of the tasks performed, most modern data-driven approaches accomplish both SD and T-F analysis, as highlighted in the table. However, some techniques are known to to better at a particular task: for instance, while SST is considered as state-of-the-art in T-F analysis, it exhibits weakness in SD owing to the lack of a robust ridge extraction technique. EMD and VMD are powerful signal decomposition approaches with T-F analysis capability coming from the application of the Hilbert transform on the decomposed components. VNCMD inherently obtains signal decomposition as well as T-F signatures of those components while SSA has been mainly designed as a signal decomposition method.    

Regarding the accuracy of SD and T-F analysis, SST and VNCMD were found to be the best in our experiments. This was mainly due to ability of these approaches to cater for both narrow- and wide-band signals. The experiments in section 3 confirm our observation. For narrow-band signals, VMD was found to be superior as compared to both EMD and SSA. It should also be noted that only SST comes with theoretical guarantees for accurate decomposition of signals in terms of AM-FM components. Like their univariate counterparts, MVMD is overall superior to MEMD in terms of avoiding the mode-mixing problem. Moreover, as illustrated in the Figure 8, MVMD better aligns similar frequency components across channels (mode-alignment) as compared to MEMD.  

In terms of noise robustness, SST and VMD performed comparatively well. Note that among all the methods that are considered here, SST is the only one that is complemented with theoretical guarantees for noise robustness \cite{Thakur13}. While the VNCMD performed exceedingly well in some cases, we found that its performance was quite erratic on other noise realizations; so much so that in some cases, the VNCMD even failed to converge. The performance of EMD and SSA methods deteriorated even in the presence of small amounts of noise. The multivariate extensions of EMD and VMD inherit the noise robustness properties of their univariate counterparts i.e., MVMD is robust to noise as compared to MEMD.  

The VNCMD is highly sensitive to changes in its parameters and was found to be difficult to tune in all our experiments. Particularly, it required very accurate initial estimates of center frequencies to produce any meaningful results. The performance of VMD and EMD was generally stable to changes in their parameters. The EMD is the only method that did not require the number of components $K$ to be defined a priori. The SST was also robust to changes in parameters but required proper tweaking of its ridge extraction scheme to deliver meaningful SD results. The SSA was found to be very sensitive to changes in its parameters. Finally, like their univariate counterparts, the multivariate extensions of EMD and VMD were found to be highly and moderately robust to paramater changes respectively.   

While a detailed analysis of the computational complexity of the data-driven SD approaches is beyond the scope of this work, we comment on the complexity of the SD methods based on our experimental observations on the ensemble of $s_1$ and $s_2$. Only the time complexity was considered here and not the space/memory complexity. We found VMD to be the fastest of all the methods followed by EMD, SSA, VNCMD and SST. Among the multivariate extensions, the MVMD was found to be more computationally efficient than MEMD.  

\section{Conclusion}
A comparative analysis of modern data-driven signal decomposition and time-frequency methods has been performed. The methods include empirical mode decomposition (EMD), variational mode decomposition (VMD), variational nonlinear chirp mode decomposition (VNCMD), synchrosqueezed transform (SST) and sliding singular spectrum analysis (SSA). In addition, we have also compared the performance of the two popular multivariate extensions of the data-driven signal decomposition methods, including multivariate VMD (MVMD) and multivariate EMD (MEMD). The methods have been evaluated in terms of their accuracy to decompose non-stationary signals into AM-FM like components, robustness to noise, robustness to changes in their parameters and alignment of similar frequency components along multiple channels of a multivariate signal (mode-alignment). This has been achieved via multiple experiments involving carefully designed synthetic signals (with narrow- and wide-band properties) and real-life biomedical signals. Our observations from those experiments have been summarized in Table 1 which show that the SST performs best in terms of accuracy and robustness to noise, whereas VMD is superior in respect of robustness to parameter changes and speed while also being reasonably accurate and robust to noise. The performance of the SSA and EMD methods has been found to be relatively sub-optimal both in terms of accuracy and robustness to noise. Finally, it has been observed that VNCMD is quite erratic in its performance, performing well in some cases while not converging at all in others -- in addition to the fact that it is highly difficult to tune its parameters to yield reasonable performance. Among multivariate approaches, we have observed MVMD to be superior to MEMD in terms of accuracy, noise robustness, mode-alignment and computational efficiency.   

For future work in this area, the following avenues/directions could be considered: i) no available signal decomposition technique is adequately robust to noise under simple conditions, calling for novel robust SD approaches; ii) no existing SD technique provides complete guarantees for accuracy/correctness and robustness to noise under practical conditions; and iii) all existing data-driven techniques are computationally expensive (when compared against wavelets and short-time Fourier transform), especially for medium- to large-sized (e.g., multivariate) data.

\section{Data and Code Availability}
The synthetic data sets analysed during this study can be easily generated by using the signal models given in the manuscript. The real-life cardiotocographic data used in section 6 is freely available  here \cite{Physionet} while the EEG data used in section 3 can be made available by the corresponding author on request. The following freely available codes of the popular SD methods were used in the study:\\
EMD: https://perso.ens-lyon.fr/patrick.flandrin/emd.html\\
MEMD: https://www.commsp.ee.ic.ac.uk/~mandic/research/emd.htm\\
VMD: https://se.mathworks.com/matlabcentral/fileexchange/44765-variational-mode-decomposition\\
MVMD: https://se.mathworks.com/matlabcentral/fileexchange/72814-multivariate-variational-mode-decomposition-mvmd\\
SST: Built-in function within MATLAB (\textit{wsst})\\
SSA: https://codeocean.com/capsule/5444528/tree/v1\\
VNCMD: https://se.mathworks.com/matlabcentral/fileexchange/64292-variational-nonlinear-chirp-mode-decomposition\\
\bibliography{References}

\begin{thebibliography}{10}
\urlstyle{rm}
\expandafter\ifx\csname url\endcsname\relax
  \def\url#1{\texttt{#1}}\fi
\expandafter\ifx\csname urlprefix\endcsname\relax\def\urlprefix{URL }\fi
\expandafter\ifx\csname doiprefix\endcsname\relax\def\doiprefix{DOI: }\fi
\providecommand{\bibinfo}[2]{#2}
\providecommand{\eprint}[2][]{\url{#2}}

\bibitem{Flandrin18}
\bibinfo{author}{Flandrin, P.}
\newblock \emph{\bibinfo{title}{Explorations in time-frequency analysis}}
  (\bibinfo{publisher}{Cambridge University Press},
  \bibinfo{address}{Cambridge, England}, \bibinfo{year}{2018}).

\bibitem{Daub11}
\bibinfo{author}{Daubechies, I.}, \bibinfo{author}{Lu, J.} \&
  \bibinfo{author}{Wu, H.-T.}
\newblock \bibinfo{journal}{\bibinfo{title}{Synchrosqueezed wavelet transforms:
  An empirical mode decomposition-like tool}}.
\newblock {\emph{\JournalTitle{Applied and Computational Harmonic Analysis}}}
  \textbf{\bibinfo{volume}{30}}, \bibinfo{pages}{243--261}
  (\bibinfo{year}{2011}).

\bibitem{Takeda11}
\bibinfo{author}{Takeda, N.} \& \bibinfo{author}{Maemura, K.}
\newblock \bibinfo{journal}{\bibinfo{title}{Circadian clock and cardiovascular
  disease}}.
\newblock {\emph{\JournalTitle{Journal of Cardiology}}}
  \textbf{\bibinfo{volume}{57}}, \bibinfo{pages}{249--256}
  (\bibinfo{year}{2011}).

\bibitem{Wang10}
\bibinfo{author}{Wang, X.-J.}
\newblock \bibinfo{journal}{\bibinfo{title}{Neurophysiological and
  computational principles of cortical rhythms in cognition}}.
\newblock {\emph{\JournalTitle{Physiological Reviews}}}
  \textbf{\bibinfo{volume}{90}}, \bibinfo{pages}{1195--1268}
  (\bibinfo{year}{2010}).

\bibitem{Lin14}
\bibinfo{author}{Lin, Y.-T.}, \bibinfo{author}{Wu, H.-T.},
  \bibinfo{author}{Tsao, J.}, \bibinfo{author}{Yien, H.-W.} \&
  \bibinfo{author}{Hseu, S.-S.}
\newblock \bibinfo{journal}{\bibinfo{title}{Time-varying spectral analysis
  revealing differential effects of sevoflurane anaesthesia:
  non-rhythmic-to-rhythmic ratio}}.
\newblock {\emph{\JournalTitle{Acta Anaesthesiologica Scandinavica}}}
  \textbf{\bibinfo{volume}{58}}, \bibinfo{pages}{157--167}
  (\bibinfo{year}{2014}).

\bibitem{Baudin14}
\bibinfo{author}{Baudin, F.} \emph{et~al.}
\newblock \bibinfo{journal}{\bibinfo{title}{Impact of ventilatory modes on the
  breathing variability in mechanically ventilated infants}}.
\newblock {\emph{\JournalTitle{Front. Pediatr.}}} \textbf{\bibinfo{volume}{2}},
  \bibinfo{pages}{132} (\bibinfo{year}{2014}).

\bibitem{Dybala14}
\bibinfo{author}{Dyba{\l}a, J.} \& \bibinfo{author}{Zimroz, R.}
\newblock \bibinfo{journal}{\bibinfo{title}{Rolling bearing diagnosing method
  based on empirical mode decomposition of machine vibration signal}}.
\newblock {\emph{\JournalTitle{Applied Acoustics}}}
  \textbf{\bibinfo{volume}{77}}, \bibinfo{pages}{195--203}
  (\bibinfo{year}{2014}).

\bibitem{Maragos01}
\bibinfo{author}{Potamianos, A.} \& \bibinfo{author}{Maragos, P.}
\newblock \bibinfo{journal}{\bibinfo{title}{Time-frequency distributions for
  automatic speech recognition}}.
\newblock {\emph{\JournalTitle{IEEE Transactions on Speech and Audio
  Processing}}} \textbf{\bibinfo{volume}{9}}, \bibinfo{pages}{196--200}
  (\bibinfo{year}{2001}).

\bibitem{Mallat08}
\bibinfo{author}{Yu, G.}, \bibinfo{author}{Mallat, S.} \&
  \bibinfo{author}{Bacry, E.}
\newblock \bibinfo{journal}{\bibinfo{title}{Audio denoising by time-frequency
  block thresholding}}.
\newblock {\emph{\JournalTitle{IEEE Transactions on Signal Processing}}}
  \textbf{\bibinfo{volume}{56}}, \bibinfo{pages}{1830--1839}
  (\bibinfo{year}{2008}).

\bibitem{Matz13}
\bibinfo{author}{Matz, G.}, \bibinfo{author}{Bolcskei, H.} \&
  \bibinfo{author}{Hlawatsch, F.}
\newblock \bibinfo{journal}{\bibinfo{title}{Time-frequency foundations of
  communications: Concepts and tools}}.
\newblock {\emph{\JournalTitle{IEEE Signal Processing Magazine}}}
  \textbf{\bibinfo{volume}{30}}, \bibinfo{pages}{87--96}
  (\bibinfo{year}{2013}).

\bibitem{Park13}
\bibinfo{author}{Park, C.}, \bibinfo{author}{Looney, D.},
  \bibinfo{author}{Rehman, N.}, \bibinfo{author}{Ahrabian, A.} \&
  \bibinfo{author}{Mandic, D.~P.}
\newblock \bibinfo{journal}{\bibinfo{title}{Classification of motor imagery bci
  using multivariate empirical mode decomposition}}.
\newblock {\emph{\JournalTitle{IEEE Transactions on Neural Systems and
  Rehabilitation Engineering}}} \textbf{\bibinfo{volume}{21}},
  \bibinfo{pages}{10--22} (\bibinfo{year}{2013}).

\bibitem{Saleem19}
\bibinfo{author}{Saleem, S.} \emph{et~al.}
\newblock \bibinfo{journal}{\bibinfo{title}{A strategy for classification of
  ``vaginal vs. cesarean section'' delivery: Bivariate empirical mode
  decomposition of cardiotocographic recordings}}.
\newblock {\emph{\JournalTitle{Front. Physiol.}}}
  \textbf{\bibinfo{volume}{10}}, \bibinfo{pages}{246} (\bibinfo{year}{2019}).

\bibitem{Rehman17}
\bibinfo{author}{Rehman, N.} \emph{et~al.}
\newblock \bibinfo{journal}{\bibinfo{title}{Translation invariant multi-scale
  signal denoising based on goodness-of-fit tests}}.
\newblock {\emph{\JournalTitle{Signal Processing}}}
  \textbf{\bibinfo{volume}{131}}, \bibinfo{pages}{220--234}
  (\bibinfo{year}{2017}).

\bibitem{Rehman19-den}
\bibinfo{author}{Rehman, N.}, \bibinfo{author}{Khan, B.} \&
  \bibinfo{author}{Naveed, K.}
\newblock \bibinfo{journal}{\bibinfo{title}{Data-driven multivariate signal
  denoising using mahalanobis distance}}.
\newblock {\emph{\JournalTitle{IEEE Signal Processing Letters}}}
  \textbf{\bibinfo{volume}{26}}, \bibinfo{pages}{1408--1412}
  (\bibinfo{year}{2019}).

\bibitem{Naveed20}
\bibinfo{author}{Naveed, K.} \& \bibinfo{author}{Rehman, N.}
\newblock \bibinfo{journal}{\bibinfo{title}{Wavelet based multivariate signal
  denoising using mahalanobis distance and edf statistics}}.
\newblock {\emph{\JournalTitle{IEEE Transactions on Signal Processing}}}
  \textbf{\bibinfo{volume}{68}}, \bibinfo{pages}{5997--6010}
  (\bibinfo{year}{2020}).

\bibitem{Naveed21}
\bibinfo{author}{Naveed, K.}, \bibinfo{author}{Akhtar, M.~T.},
  \bibinfo{author}{Siddiqui, M.~F.} \& \bibinfo{author}{Rehman, N.}
\newblock \bibinfo{journal}{\bibinfo{title}{A statistical approach to signal
  denoising based on data-driven multiscale representation}}.
\newblock {\emph{\JournalTitle{Digital Signal Processing}}}
  \textbf{\bibinfo{volume}{108}}, \bibinfo{pages}{102896}
  (\bibinfo{year}{2021}).

\bibitem{Rehman09-fus}
\bibinfo{author}{Rehman, N.}, \bibinfo{author}{Looney, D.} \&
  \bibinfo{author}{Mandic, D.}
\newblock \bibinfo{title}{{Bivariate EMD-based image fusion}}.
\newblock In \emph{\bibinfo{booktitle}{{Proceedings of the 15th IEEE/SP
  Workshop on Statistical Signal Processing}}} (\bibinfo{address}{Cardiff,
  Wales}, \bibinfo{year}{2009}).

\bibitem{Rehman15}
\bibinfo{author}{Rehman, N.} \emph{et~al.}
\newblock \bibinfo{journal}{\bibinfo{title}{Multi-scale pixel-based image
  fusion using multivariate empirical mode decomposition}}.
\newblock {\emph{\JournalTitle{Sensors}}} \textbf{\bibinfo{volume}{15}},
  \bibinfo{pages}{10923--10947} (\bibinfo{year}{2015}).

\bibitem{Abd15}
\bibinfo{author}{Abdullah, S. M.~U.}, \bibinfo{author}{Rehman, N.},
  \bibinfo{author}{Khan, M.~M.} \& \bibinfo{author}{Mandic, D.~P.}
\newblock \bibinfo{journal}{\bibinfo{title}{A multivariate empirical mode
  decomposition based approach to pansharpening}}.
\newblock {\emph{\JournalTitle{IEEE Transactions on Geoscience and Remote
  Sensing}}} \textbf{\bibinfo{volume}{53}}, \bibinfo{pages}{3974--3984}
  (\bibinfo{year}{2015}).

\bibitem{Daub14}
\bibinfo{author}{Wu, H.-T.}, \bibinfo{author}{Hseu, S.-S.},
  \bibinfo{author}{Bien, M.-Y.}, \bibinfo{author}{Kou, Y.~R.} \&
  \bibinfo{author}{Daubechies, I.}
\newblock \bibinfo{journal}{\bibinfo{title}{Evaluating physiological dynamics
  via synchrosqueezing: prediction of ventilator weaning}}.
\newblock {\emph{\JournalTitle{IEEE Trans. Biomed. Eng.}}}
  \textbf{\bibinfo{volume}{61}}, \bibinfo{pages}{736--744}
  (\bibinfo{year}{2014}).

\bibitem{Huang98}
\bibinfo{author}{Huang, N.} \emph{et~al.}
\newblock \bibinfo{journal}{\bibinfo{title}{The empirical mode decomposition
  and the hilbert spectrum for nonlinear and non-stationary time series
  analysis}}.
\newblock {\emph{\JournalTitle{Proceedings of the Royal Society of London.
  Series A: Mathematical, Physical and Engineering Sciences}}}
  \textbf{\bibinfo{volume}{454}}, \bibinfo{pages}{903--995}
  (\bibinfo{year}{1998}).

\bibitem{Thakur13}
\bibinfo{author}{Thakur, G.}, \bibinfo{author}{Brevdo, E.},
  \bibinfo{author}{Fučkar, N.~S.} \& \bibinfo{author}{Wu, H.-T.}
\newblock \bibinfo{journal}{\bibinfo{title}{The synchrosqueezing algorithm for
  time-varying spectral analysis: Robustness properties and new paleoclimate
  applications}}.
\newblock {\emph{\JournalTitle{Signal Processing}}}
  \textbf{\bibinfo{volume}{93}}, \bibinfo{pages}{1079--1094}
  (\bibinfo{year}{2013}).

\bibitem{VMD14}
\bibinfo{author}{Dragomiretskiy, K.} \& \bibinfo{author}{Zosso, D.}
\newblock \bibinfo{journal}{\bibinfo{title}{Variational mode decomposition}}.
\newblock {\emph{\JournalTitle{IEEE Transactions on Signal Processing}}}
  \textbf{\bibinfo{volume}{62}}, \bibinfo{pages}{531--544}
  (\bibinfo{year}{2014}).

\bibitem{NCMD18}
\bibinfo{author}{Chen, S.}, \bibinfo{author}{Dong, X.}, \bibinfo{author}{Peng,
  Z.}, \bibinfo{author}{Zhang, W.} \& \bibinfo{author}{Meng, G.}
\newblock \bibinfo{journal}{\bibinfo{title}{Nonlinear chirp mode decomposition:
  A variational method}}.
\newblock {\emph{\JournalTitle{IEEE Transactions on Signal Processing}}}
  \textbf{\bibinfo{volume}{65}}, \bibinfo{pages}{6024--6037}
  (\bibinfo{year}{2017}).

\bibitem{Fland18}
\bibinfo{author}{Harmouche, J.}, \bibinfo{author}{Fourer, D.},
  \bibinfo{author}{Auger, F.}, \bibinfo{author}{Borgnat, P.} \&
  \bibinfo{author}{Flandrin, P.}
\newblock \bibinfo{journal}{\bibinfo{title}{{The Sliding Singular Spectrum
  Analysis: a Data-Driven Non-Stationary Signal Decomposition Tool}}}.
\newblock {\emph{\JournalTitle{{IEEE Transactions on Signal Processing}}}}
  (\bibinfo{year}{2017}).

\bibitem{Huang09}
\bibinfo{author}{Huang, N.~E.} \emph{et~al.}
\newblock \bibinfo{journal}{\bibinfo{title}{On instantaneous frequency}}.
\newblock {\emph{\JournalTitle{Advances in Adaptive Data Analysis}}}
  \textbf{\bibinfo{volume}{01}}, \bibinfo{pages}{177--229}
  (\bibinfo{year}{2009}).

\bibitem{Meignen07}
\bibinfo{author}{Meignen, S.} \& \bibinfo{author}{Perrier, V.}
\newblock \bibinfo{journal}{\bibinfo{title}{A new formulation for empirical
  mode decomposition based on constrained optimization}}.
\newblock {\emph{\JournalTitle{{IEEE} Signal Processing Letters}}}
  \textbf{\bibinfo{volume}{14}}, \bibinfo{pages}{932--935}
  (\bibinfo{year}{2007}).

\bibitem{Pustelnik14}
\bibinfo{author}{Pustelnik, N.}, \bibinfo{author}{Borgnat, P.} \&
  \bibinfo{author}{Flandrin, P.}
\newblock \bibinfo{journal}{\bibinfo{title}{Empirical mode decomposition
  revisited by multicomponent non-smooth convex optimization}}.
\newblock {\emph{\JournalTitle{Signal Processing}}}
  \textbf{\bibinfo{volume}{102}}, \bibinfo{pages}{313--331}
  (\bibinfo{year}{2014}).

\bibitem{Lin09}
\bibinfo{author}{Lin, L.}, \bibinfo{author}{Wang, Y.} \& \bibinfo{author}{Zhou,
  H.}
\newblock \bibinfo{journal}{\bibinfo{title}{Iterative filtering as an
  alternative algorithm for empirical mode decomposition}}.
\newblock {\emph{\JournalTitle{Advances in Adaptive Data Analysis}}}
  \textbf{\bibinfo{volume}{01}}, \bibinfo{pages}{543--560}
  (\bibinfo{year}{2009}).

\bibitem{Cicone16}
\bibinfo{author}{Cicone, A.}, \bibinfo{author}{Liu, J.} \&
  \bibinfo{author}{Zhou, H.}
\newblock \bibinfo{journal}{\bibinfo{title}{Adaptive local iterative filtering
  for signal decomposition and instantaneous frequency analysis}}.
\newblock {\emph{\JournalTitle{Applied and Computational Harmonic Analysis}}}
  \textbf{\bibinfo{volume}{41}}, \bibinfo{pages}{384--411}
  (\bibinfo{year}{2016}).

\bibitem{Hou11}
\bibinfo{author}{Hou, T.} \& \bibinfo{author}{Shi, Z.}
\newblock \bibinfo{journal}{\bibinfo{title}{Adaptive data analysis via sparse
  time-frequency representation.}}
\newblock {\emph{\JournalTitle{Advances in Adaptive Data Analysis}}}
  \textbf{\bibinfo{volume}{3}}, \bibinfo{pages}{1--28} (\bibinfo{year}{2011}).

\bibitem{Hou16}
\bibinfo{author}{Hou, T.~Y.} \& \bibinfo{author}{Shi, Z.}
\newblock \bibinfo{journal}{\bibinfo{title}{Sparse time-frequency decomposition
  based on dictionary adaptation}}.
\newblock {\emph{\JournalTitle{Philosophical Transactions of the Royal Society
  A: Mathematical, Physical and Engineering Sciences}}}
  \textbf{\bibinfo{volume}{374}}, \bibinfo{pages}{20150192}
  (\bibinfo{year}{2016}).

\bibitem{Wu09}
\bibinfo{author}{Wu, Z.} \& \bibinfo{author}{Huang, N.}
\newblock \bibinfo{journal}{\bibinfo{title}{Ensemble empirical mode
  decomposition: a noise-assisted data analysis method}}.
\newblock {\emph{\JournalTitle{Advances in Adaptive Data Analysis}}}
  \textbf{\bibinfo{volume}{1}}, \bibinfo{pages}{1--41} (\bibinfo{year}{2009}).

\bibitem{Rehman13-AADA}
\bibinfo{author}{Rehman, N.}, \bibinfo{author}{Park, C.},
  \bibinfo{author}{Huang, N.} \& \bibinfo{author}{Mandic, D.}
\newblock \bibinfo{journal}{\bibinfo{title}{Emd via memd: multivariate
  noise-aided computation of standard emd}}.
\newblock {\emph{\JournalTitle{Advances in Adaptive Data Analysis}}}
  \textbf{\bibinfo{volume}{05}}, \bibinfo{pages}{1–25}
  (\bibinfo{year}{2013}).

\bibitem{Lang20}
\bibinfo{author}{Lang, X.}, \bibinfo{author}{Rehman, N.},
  \bibinfo{author}{Zhang, Y.}, \bibinfo{author}{Xie, L.} \&
  \bibinfo{author}{Su, H.}
\newblock \bibinfo{journal}{\bibinfo{title}{Median ensemble empirical mode
  decomposition}}.
\newblock {\emph{\JournalTitle{Signal Processing}}}
  \textbf{\bibinfo{volume}{176}}, \bibinfo{pages}{107686}
  (\bibinfo{year}{2020}).

\bibitem{EMD-alg}
\bibinfo{author}{Rilling, G.}, \bibinfo{author}{Flandrin, P.} \&
  \bibinfo{author}{Gon{\c c}alves, P.}
\newblock \bibinfo{title}{{On empirical mode decomposition and its
  algorithms}}.
\newblock In \emph{\bibinfo{booktitle}{{Proceedings of IEEE-EURASIP Workshop on
  Nonlinear Signal and Image Processing NSIP-03}}} (\bibinfo{address}{Grado,
  Italy}, \bibinfo{year}{2003}).

\bibitem{Daub96}
\bibinfo{author}{Daubechies, I.} \& \bibinfo{author}{Maes, S.}
\newblock \bibinfo{title}{A nonlinear squeezing of the continuous wavelet
  transform based on auditory nerve models}.
\newblock In \bibinfo{editor}{Aldroubi, A.} \& \bibinfo{editor}{Unser, M.}
  (eds.) \emph{\bibinfo{booktitle}{Wavelets in medicine and biology}},
  chap.~\bibinfo{chapter}{20}, \bibinfo{pages}{527--546}
  (\bibinfo{publisher}{CRC Press}, \bibinfo{address}{Boca Raton, FL},
  \bibinfo{year}{1996}).
\newblock \bibinfo{note}{Zbl:0848.92003.}

\bibitem{SST_SPM13}
\bibinfo{author}{Auger, F.} \emph{et~al.}
\newblock \bibinfo{journal}{\bibinfo{title}{Time-frequency reassignment and
  synchrosqueezing: An overview}}.
\newblock {\emph{\JournalTitle{IEEE Signal Processing Magazine}}}
  \textbf{\bibinfo{volume}{30}}, \bibinfo{pages}{32--41}
  (\bibinfo{year}{2013}).

\bibitem{Thakur11}
\bibinfo{author}{Thakur, G.} \& \bibinfo{author}{Wu, H.-T.}
\newblock \bibinfo{journal}{\bibinfo{title}{Synchrosqueezing-based recovery of
  instantaneous frequency from nonuniform samples}}.
\newblock {\emph{\JournalTitle{{SIAM} Journal on Mathematical Analysis}}}
  \textbf{\bibinfo{volume}{43}}, \bibinfo{pages}{2078--2095}
  (\bibinfo{year}{2011}).

\bibitem{Huang16}
\bibinfo{author}{lai Huang, Z.}, \bibinfo{author}{Zhang, J.},
  \bibinfo{author}{hu~Zhao, T.} \& \bibinfo{author}{Sun, Y.}
\newblock \bibinfo{journal}{\bibinfo{title}{Synchrosqueezing s-transform and
  its application in seismic spectral decomposition}}.
\newblock {\emph{\JournalTitle{{IEEE} Transactions on Geoscience and Remote
  Sensing}}} \textbf{\bibinfo{volume}{54}}, \bibinfo{pages}{817--825}
  (\bibinfo{year}{2016}).

\bibitem{daub15}
\bibinfo{author}{Daubechies, I.}, \bibinfo{author}{Wang, Y.~G.} \&
  \bibinfo{author}{tieng Wu, H.}
\newblock \bibinfo{journal}{\bibinfo{title}{{ConceFT}: concentration of
  frequency and time via a multitapered synchrosqueezed transform}}.
\newblock {\emph{\JournalTitle{Philosophical Transactions of the Royal Society
  A: Mathematical, Physical and Engineering Sciences}}}
  \textbf{\bibinfo{volume}{374}}, \bibinfo{pages}{20150193}
  (\bibinfo{year}{2016}).

\bibitem{Oberlin15}
\bibinfo{author}{Oberlin, T.}, \bibinfo{author}{Meignen, S.} \&
  \bibinfo{author}{Perrier, V.}
\newblock \bibinfo{journal}{\bibinfo{title}{Second-order synchrosqueezing
  transform or invertible reassignment? towards ideal time-frequency
  representations}}.
\newblock {\emph{\JournalTitle{{IEEE} Transactions on Signal Processing}}}
  \textbf{\bibinfo{volume}{63}}, \bibinfo{pages}{1335--1344}
  (\bibinfo{year}{2015}).

\bibitem{Bahera18}
\bibinfo{author}{Behera, R.}, \bibinfo{author}{Meignen, S.} \&
  \bibinfo{author}{Oberlin, T.}
\newblock \bibinfo{journal}{\bibinfo{title}{Theoretical analysis of the
  second-order synchrosqueezing transform}}.
\newblock {\emph{\JournalTitle{Applied and Computational Harmonic Analysis}}}
  \textbf{\bibinfo{volume}{45}}, \bibinfo{pages}{379--404}
  (\bibinfo{year}{2018}).

\bibitem{Pham17}
\bibinfo{author}{Pham, D.-H.} \& \bibinfo{author}{Meignen, S.}
\newblock \bibinfo{journal}{\bibinfo{title}{High-order synchrosqueezing
  transform for multicomponent signals analysis—with an application to
  gravitational-wave signal}}.
\newblock {\emph{\JournalTitle{IEEE Transactions on Signal Processing}}}
  \textbf{\bibinfo{volume}{65}}, \bibinfo{pages}{3168--3178}
  (\bibinfo{year}{2017}).

\bibitem{Yang18}
\bibinfo{author}{Yang, H.}
\newblock \bibinfo{journal}{\bibinfo{title}{Statistical analysis of
  synchrosqueezed transforms}}.
\newblock {\emph{\JournalTitle{Applied and Computational Harmonic Analysis}}}
  \textbf{\bibinfo{volume}{45}}, \bibinfo{pages}{526--550}
  (\bibinfo{year}{2018}).

\bibitem{Chui16}
\bibinfo{author}{Chui, C.~K.} \& \bibinfo{author}{Mhaskar, H.}
\newblock \bibinfo{journal}{\bibinfo{title}{Signal decomposition and analysis
  via extraction of frequencies}}.
\newblock {\emph{\JournalTitle{Applied and Computational Harmonic Analysis}}}
  \textbf{\bibinfo{volume}{40}}, \bibinfo{pages}{97--136}
  (\bibinfo{year}{2016}).

\bibitem{Li20}
\bibinfo{author}{Li, L.}, \bibinfo{author}{Chui, C.~K.} \&
  \bibinfo{author}{Jiang, Q.}
\newblock \bibinfo{title}{Direct signal separation via extraction of local
  frequencies with adaptive time-varying parameters} (\bibinfo{year}{2020}).

\bibitem{Meignen17}
\bibinfo{author}{Meignen, S.}, \bibinfo{author}{Pham, D.-H.} \&
  \bibinfo{author}{McLaughlin, S.}
\newblock \bibinfo{journal}{\bibinfo{title}{On demodulation, ridge detection,
  and synchrosqueezing for multicomponent signals}}.
\newblock {\emph{\JournalTitle{{IEEE} Transactions on Signal Processing}}}
  \textbf{\bibinfo{volume}{65}}, \bibinfo{pages}{2093--2103}
  (\bibinfo{year}{2017}).

\bibitem{Laurent20}
\bibinfo{author}{Laurent, N.} \& \bibinfo{author}{Meignen, S.}
\newblock \bibinfo{journal}{\bibinfo{title}{A novel ridge detector for
  nonstationary multicomponent signals: Development and application to robust
  mode retrieval}}.
\newblock {\emph{\JournalTitle{IEEE Transactions on Signal Processing}}}
  \textbf{\bibinfo{volume}{69}}, \bibinfo{pages}{3325--3336}
  (\bibinfo{year}{2021}).

\bibitem{Comon94}
\bibinfo{author}{Comon, P.}
\newblock \bibinfo{journal}{\bibinfo{title}{Independent component analysis, a
  new concept?}}
\newblock {\emph{\JournalTitle{Signal Processing}}}
  \textbf{\bibinfo{volume}{36}}, \bibinfo{pages}{287--314}
  (\bibinfo{year}{1994}).

\bibitem{Lee99}
\bibinfo{author}{Lee, D.~D.} \& \bibinfo{author}{Seung, H.~S.}
\newblock \bibinfo{journal}{\bibinfo{title}{Learning the parts of objects by
  non-negative matrix factorization}}.
\newblock {\emph{\JournalTitle{Nature}}} \textbf{\bibinfo{volume}{401}},
  \bibinfo{pages}{788--791} (\bibinfo{year}{1999}).

\bibitem{Smaragdis14}
\bibinfo{author}{Smaragdis, P.}, \bibinfo{author}{Fevotte, C.},
  \bibinfo{author}{Mysore, G.~J.}, \bibinfo{author}{Mohammadiha, N.} \&
  \bibinfo{author}{Hoffman, M.}
\newblock \bibinfo{journal}{\bibinfo{title}{Static and dynamic source
  separation using nonnegative factorizations: A unified view}}.
\newblock {\emph{\JournalTitle{{IEEE} Signal Processing Magazine}}}
  \textbf{\bibinfo{volume}{31}}, \bibinfo{pages}{66--75}
  (\bibinfo{year}{2014}).

\bibitem{Broomhead86}
\bibinfo{author}{Broomhead, D.} \& \bibinfo{author}{King, G.~P.}
\newblock \bibinfo{journal}{\bibinfo{title}{Extracting qualitative dynamics
  from experimental data}}.
\newblock {\emph{\JournalTitle{Physica D: Nonlinear Phenomena}}}
  \textbf{\bibinfo{volume}{20}}, \bibinfo{pages}{217--236}
  (\bibinfo{year}{1986}).

\bibitem{LTFRS18}
\bibinfo{author}{Fevotte, C.} \& \bibinfo{author}{Kowalski, M.}
\newblock \bibinfo{journal}{\bibinfo{title}{Estimation with low-rank
  time{\textendash}frequency synthesis models}}.
\newblock {\emph{\JournalTitle{{IEEE} Transactions on Signal Processing}}}
  \textbf{\bibinfo{volume}{66}}, \bibinfo{pages}{4121--4132}
  (\bibinfo{year}{2018}).

\bibitem{Rehman09}
\bibinfo{author}{Rehman, N.} \& \bibinfo{author}{Mandic, D.~P.}
\newblock \bibinfo{journal}{\bibinfo{title}{Multivariate empirical mode
  decomposition}}.
\newblock {\emph{\JournalTitle{Proceedings of the Royal Society A:
  Mathematical, Physical and Engineering Sciences}}}
  \textbf{\bibinfo{volume}{466}}, \bibinfo{pages}{1291--1302}
  (\bibinfo{year}{2009}).

\bibitem{Rehman10}
\bibinfo{author}{Rehman, N.} \& \bibinfo{author}{Mandic, D.~P.}
\newblock \bibinfo{journal}{\bibinfo{title}{Empirical mode decomposition for
  trivariate signals}}.
\newblock {\emph{\JournalTitle{{IEEE} Transactions on Signal Processing}}}
  \textbf{\bibinfo{volume}{58}}, \bibinfo{pages}{1059--1068}
  (\bibinfo{year}{2010}).

\bibitem{Rehman11}
\bibinfo{author}{Rehman, N.} \& \bibinfo{author}{Mandic, D.~P.}
\newblock \bibinfo{journal}{\bibinfo{title}{Filter bank property of
  multivariate empirical mode decomposition}}.
\newblock {\emph{\JournalTitle{{IEEE} Transactions on Signal Processing}}}
  \textbf{\bibinfo{volume}{59}}, \bibinfo{pages}{2421--2426}
  (\bibinfo{year}{2011}).

\bibitem{Rehman13}
\bibinfo{author}{Mandic, D.~P.}, \bibinfo{author}{Rehman, N.},
  \bibinfo{author}{Wu, Z.} \& \bibinfo{author}{Huang, N.~E.}
\newblock \bibinfo{journal}{\bibinfo{title}{Empirical mode decomposition-based
  time-frequency analysis of multivariate signals: The power of adaptive data
  analysis}}.
\newblock {\emph{\JournalTitle{{IEEE} Signal Processing Magazine}}}
  \textbf{\bibinfo{volume}{30}}, \bibinfo{pages}{74--86}
  (\bibinfo{year}{2013}).

\bibitem{Cicone19}
\bibinfo{author}{Cicone, A.} \& \bibinfo{author}{Pellegrino, E.}
\newblock \bibinfo{title}{Multivariate fast iterative filtering for the
  decomposition of nonstationary signals} (\bibinfo{year}{2019}).

\bibitem{Rehman19}
\bibinfo{author}{Rehman, N.} \& \bibinfo{author}{Aftab, H.}
\newblock \bibinfo{journal}{\bibinfo{title}{Multivariate variational mode
  decomposition}}.
\newblock {\emph{\JournalTitle{IEEE Transactions on Signal Processing}}}
  \textbf{\bibinfo{volume}{67}}, \bibinfo{pages}{6039--6052}
  (\bibinfo{year}{2019}).

\bibitem{Chen20}
\bibinfo{author}{Chen, Q.}, \bibinfo{author}{Xie, L.} \& \bibinfo{author}{Su,
  H.}
\newblock \bibinfo{journal}{\bibinfo{title}{Multivariate nonlinear chirp mode
  decomposition}}.
\newblock {\emph{\JournalTitle{Signal Processing}}}
  \textbf{\bibinfo{volume}{176}}, \bibinfo{pages}{107667}
  (\bibinfo{year}{2020}).

\bibitem{Ali15}
\bibinfo{author}{Ahrabian, A.}, \bibinfo{author}{Looney, D.},
  \bibinfo{author}{Stanković, L.} \& \bibinfo{author}{Mandic, D.~P.}
\newblock \bibinfo{journal}{\bibinfo{title}{Synchrosqueezing-based
  time-frequency analysis of multivariate data}}.
\newblock {\emph{\JournalTitle{Signal Processing}}}
  \textbf{\bibinfo{volume}{106}}, \bibinfo{pages}{331--341}
  (\bibinfo{year}{2015}).

\bibitem{MEWT18}
\bibinfo{author}{Singh, O.} \& \bibinfo{author}{Sunkaria, R.~K.}
\newblock \bibinfo{journal}{\bibinfo{title}{An empirical wavelet transform
  based approach for multivariate data processing application to cardiovascular
  physiological signals}}.
\newblock {\emph{\JournalTitle{Bio-Algorithms and Med-Systems}}}
  \textbf{\bibinfo{volume}{14}} (\bibinfo{year}{2018}).

\bibitem{Jain20}
\bibinfo{author}{Jain, S.}, \bibinfo{author}{Panda, R.} \&
  \bibinfo{author}{Tripathy, R.~K.}
\newblock \bibinfo{journal}{\bibinfo{title}{Multivariate sliding-mode singular
  spectrum analysis for the decomposition of multisensor time series}}.
\newblock {\emph{\JournalTitle{IEEE Sensors Letters}}}
  \textbf{\bibinfo{volume}{4}}, \bibinfo{pages}{1--4} (\bibinfo{year}{2020}).

\bibitem{Nazari19}
\bibinfo{author}{Nazari, M.} \& \bibinfo{author}{Sakhaei, S.~M.}
\newblock \bibinfo{journal}{\bibinfo{title}{Successive variational mode
  decomposition}}.
\newblock {\emph{\JournalTitle{Signal Processing}}}
  \textbf{\bibinfo{volume}{174}}, \bibinfo{pages}{107610}
  (\bibinfo{year}{2020}).

\bibitem{Qiming21}
\bibinfo{author}{Chen, Q.} \emph{et~al.}
\newblock \bibinfo{journal}{\bibinfo{title}{Self-tuning variational mode
  decomposition}}.
\newblock {\emph{\JournalTitle{Journal of the Franklin Institute}}}
  \textbf{\bibinfo{volume}{358}}, \bibinfo{pages}{7825--7862}
  (\bibinfo{year}{2021}).

\bibitem{Physionet}
\bibinfo{author}{Chudáček, V.} \emph{et~al.}
\newblock \bibinfo{title}{The ctu-uhb intrapartum cardiotocography database}.
\newblock
  \bibinfo{howpublished}{\url{https://www.physionet.org/content/ctu-uhb-ctgdb/1.0.0}}
  (\bibinfo{year}{2014}).

\end{thebibliography}

\end{document}